\documentclass[reprint, aps, nofootinbib, prb,
superscriptaddress]{revtex4-1}
\usepackage{amsmath} \usepackage{graphicx} \usepackage{dcolumn}
\usepackage{bm} 
\usepackage{amssymb}
\usepackage{pstricks}

\begin{document}

\newlength{\figurewidth}
\setlength{\figurewidth}{\columnwidth}

\newcommand{\prtl}{\partial}
\newcommand{\la}{\left\langle}
\newcommand{\ra}{\right\rangle}
\newcommand{\dla}{\la \! \! \! \la}
\newcommand{\dra}{\ra \! \! \! \ra}
\newcommand{\we}{\widetilde}
\newcommand{\smfp}{{\mbox{\scriptsize mfp}}}
\newcommand{\smp}{{\mbox{\scriptsize mp}}}
\newcommand{\sph}{{\mbox{\scriptsize ph}}}
\newcommand{\sinhom}{{\mbox{\scriptsize inhom}}}
\newcommand{\sneigh}{{\mbox{\scriptsize neigh}}}
\newcommand{\srlxn}{{\mbox{\scriptsize rlxn}}}
\newcommand{\svibr}{{\mbox{\scriptsize vibr}}}
\newcommand{\smicro}{{\mbox{\scriptsize micro}}}
\newcommand{\scoll}{{\mbox{\scriptsize coll}}}
\newcommand{\sattr}{{\mbox{\scriptsize attr}}}
\newcommand{\sth}{{\mbox{\scriptsize th}}}
\newcommand{\sauto}{{\mbox{\scriptsize auto}}}
\newcommand{\seq}{{\mbox{\scriptsize eq}}}
\newcommand{\teq}{{\mbox{\tiny eq}}}
\newcommand{\sinn}{{\mbox{\scriptsize in}}}
\newcommand{\suni}{{\mbox{\scriptsize uni}}}
\newcommand{\tin}{{\mbox{\tiny in}}}
\newcommand{\tout}{{\mbox{\tiny out}}}
\newcommand{\scr}{{\mbox{\scriptsize cr}}}
\newcommand{\tstring}{{\mbox{\tiny string}}}
\newcommand{\sperc}{{\mbox{\scriptsize perc}}}
\newcommand{\tperc}{{\mbox{\tiny perc}}}
\newcommand{\sstring}{{\mbox{\scriptsize string}}}
\newcommand{\stheor}{{\mbox{\scriptsize theor}}}
\newcommand{\sGS}{{\mbox{\scriptsize GS}}}
\newcommand{\sBP}{{\mbox{\scriptsize BP}}}
\newcommand{\sNMT}{{\mbox{\scriptsize NMT}}}
\newcommand{\sbulk}{{\mbox{\scriptsize bulk}}}
\newcommand{\tbulk}{{\mbox{\tiny bulk}}}
\newcommand{\sXtal}{{\mbox{\scriptsize Xtal}}}
\newcommand{\sliq}{{\text{\tiny liq}}}

\newcommand{\smin}{\text{min}}
\newcommand{\smax}{\text{max}}

\newcommand{\saX}{\text{\tiny aX}}
\newcommand{\slaX}{\text{l,{\tiny aX}}}

\newcommand{\svap}{{\mbox{\scriptsize vap}}}
\newcommand{\sjam}{J}
\newcommand{\Tm}{T_m}
\newcommand{\sTS}{{\mbox{\scriptsize TS}}}
\newcommand{\sDW}{{\mbox{\tiny DW}}}
\newcommand{\cN}{{\cal N}}
\newcommand{\cB}{{\cal B}}
\newcommand{\cV}{{\cal V}}
\newcommand{\cVext}{\cV_\text{ext}}
\newcommand{\br}{\bm r}
\newcommand{\be}{\bm e}
\newcommand{\cH}{{\cal H}}
\newcommand{\cHlt}{\cH_{\mbox{\scriptsize lat}}}
\newcommand{\sthermo}{{\mbox{\scriptsize thermo}}}

\newcommand{\bu}{\bm u}
\newcommand{\bk}{\bm k}
\newcommand{\bX}{\bm X}
\newcommand{\bY}{\bm Y}
\newcommand{\bA}{\bm A}
\newcommand{\bb}{\bm b}

\newcommand{\lintf}{l_\text{intf}}

\newcommand{\DV}{\delta V_{12}}
\newcommand{\sout}{{\mbox{\scriptsize out}}}
\newcommand{\dv}{\Delta v_{1 \infty}}
\newcommand{\dvin}{\Delta v_{2 \infty}}

\newcommand{\nbo}{n^{1, b}}
\newcommand{\nbi}{n^{2, b}}

\newcommand{\sint}{\text{int}}

\def\Xint#1{\mathchoice
   {\XXint\displaystyle\textstyle{#1}}%
   {\XXint\textstyle\scriptstyle{#1}}%
   {\XXint\scriptstyle\scriptscriptstyle{#1}}%
   {\XXint\scriptscriptstyle\scriptscriptstyle{#1}}%
   \!\int}
\def\XXint#1#2#3{{\setbox0=\hbox{$#1{#2#3}{\int}$}
     \vcenter{\hbox{$#2#3$}}\kern-.5\wd0}}
\def\ddashint{\Xint=}
\def\dashint{\Xint-}
\title{Pressure in the Landau-Ginzburg Functional: \\ Pascal's Law,
  Nucleation in Fluid Mixtures, a Meanfield Theory of Amphiphilic
  Action, and Interface Wetting in Glassy Liquids}

\author{Ho Yin Chan} \affiliation{Department of Physics, University of
  Houston, Houston, TX 77204-5005}

\author{Vassiliy Lubchenko} \email{vas@uh.edu} \affiliation{Department
  of Chemistry, University of Houston, Houston, TX 77204-5003}
\affiliation{Department of Physics, University of Houston, Houston, TX
  77204-5005}

\begin{abstract}

  We set up the problem of finding the transition state for phase
  nucleation in multi-component fluid mixtures, within the
  Landau-Ginzburg density functional.  We establish an expression for
  the coordinate-dependent local pressure that applies to mixtures,
  arbitrary geometries, and certain non-equilibrium
  configurations. The expression allows one to explicitly evaluate the
  pressure in spherical geometry, \`a la van der Waals. Pascal's law
  is recovered within the Landau-Ginzburg density functional theory,
  formally analogously to how conservation of energy is recovered in
  the Lagrangian formulation of mechanics. We establish proper
  boundary conditions for certain singular functional forms of the
  bulk free energy density that allow one to obtain droplet solutions
  with thick walls in essentially closed form. The hydrodynamic modes
  responsible for mixing near the interface are explicitly identified
  in the treatment; the composition at the interface is found to
  depend only weakly on the droplet size.
  Next we develop a Landau-Ginzburg treatment of the effects of
  amphiphiles on the surface tension; the amphiphilic action is seen
  as a violation of Pascal's law. We explicitly obtain the binding
  potential for the detergent at the interface and the dependence of
  the down-renormalization of the surface tension on the activity of
  the detergent.  Finally, we argue that the renormalization of the
  activation barrier for escape from long-lived structures in glassy
  liquids can be viewed as an action of uniformly seeded, randomly
  oriented amphiphilic molecules on the interface separating two
  dissimilar aperiodic structures. This renormalization is also
  considered as a ``wetting'' of the interface. The resulting
  conclusions are consistent with the random first order transition
  (RFOT) theory.
 
\end{abstract}

\date{\today}

\maketitle

\section{Motivation}

The Gibbs phase rule~\cite{LLstat, BRR} epitomizes the basic notion
that in equilibrium, the state of a macroscopic substance is fully
specified by a small set of intensive variables, such as temperature,
pressure, and the molar fractions of chemically distinct
components. As appreciated by Gibbs some 140 years ago, equilibria
with respect to particle and momentum exchange differ from each other
in a basic way: To ensure equilibrium with regard to particle exchange
between distinct phases, the chemical potentials for each individual
component in a mixture must be matched between the phases. In
contrast, mechanical equilibrium is guaranteed already by uniformity
of the full pressure; there is no need to match the {\em partial}
pressures of individual species.
Thus with each additional component in the mixture, the number of
independent variables that could be in principle used to describe an
equilibrium phase increases. Viewed alternatively, this means that the
number of distinct phases that could be in mutual equilibrium grows
with each additional component.

Situations in which some of the phases are only {\em finite} in size,
but still containing a substantial number of molecules, can be also
described rather parsimoniously with the help of equilibrium
thermodynamics, at the modest expense of introducing a free energy
cost of an interface between distinct phases. In this picture, the
chemical potentials are still spatially uniform, while the pressure
changes discontinuously across the interface, the discontinuity
scaling with the interface curvature according to the venerable
Laplace formula. The corresponding limit of an infinitely-thin
interface can be formulated in an internally-consistent fashion for
large inclusions of minority phases,~\cite{RowlinsonWidom} but becomes
ambiguous on molecular lengthscales because at its face value, a
discontinuous pressure jump implies there is an uncompensated force
acting on a subset of molecules.

In his seminal work,~\cite{vdWinterf, RowlinsonvdW} van der Waals
sought to remove this ambiguity by explicitly accounting for the free
energy cost of {\em gradual} density variations. Within a
semi-phenomenological framework equivalent to the Landau-Ginzburg
functional theory but predating it by a half-century or so, van der
Waals treats distinct phases in physical contact and their mutual
interface on the same footing.  He shows that the pressure and density
change continuously across a finite-curvature interface between two
coexisting phases.
Van der Waals's results were rediscovered and generalized in later
analyses by Cahn and Hilliard,~\cite{CahnHilliard, Cahn1959,
  CahnHilliardIII} and Hart.~\cite{PhysRev.113.412} An alternative,
quite fruitful line of work, in which the surface tension could be
directly related to detailed molecular interactions, was spurred by
pioneering work of Kirkwood and Buff,~\cite{KirkwoodBuff1949} see
reviews in Refs.~\onlinecite{RowlinsonWidom, Evans1979, Bray}.

Here we extend the van der Waals analysis to multi-component fluids
while explicitly allowing for a density dependence of the free energy
cost for spatial variations in the density itself; the latter cost
generally varies between phases. Importantly, we are also able to
include in the treatment situations that are steady-state only
locally; this is useful for deviations from equilibrium and in the
presence of chemical reactions.  The present work is partially
motivated by a challenging problem of current interest, viz., the
formation mechanism of the puzzling mesoscopic clusters of a dense
protein fluid found in some protein solutions.~\cite{GlikoJACS2005,
  Georgalis1999, doi:10.1021/jp068827o, PVL} Pan et al.~\cite{PVL}
have put forth a microscopic scenario, by which the mesoscopic
clusters consist of a spatially inhomogeneous, off-equilibrium mixture
of monomeric protein and a transient, protein-containing complex.
Already in their preliminary analysis, Pan et al.~\cite{PVL} suggest
that in the presence of protein-complex formation and decay, the
chemical potentials of the species involved become spatially
inhomogeneous even in steady state. The inhomogeneity is largely
contained within the interface and would not be captured within the
thin interface approximation, thus necessitating a treatment at least
at the van der Waals level. The reaction-diffusion equations are
non-linear and prone to numerical instabilities; numerical solutions
lack transparency that can be afforded by analytical treatment. Yet a
certain class of functional forms for the bulk free energy, i.e., a
collection of intersecting paraboloids, allows for analytical
treatment while preserving several essential features of thick
interfaces.  Hereby, the free energy surface is quadratic throughout
except at some well-defined dividing surface.  Thus, by construction,
the free energy can experience a discontinuity in the gradient and
even in the function itself. How does one ``patch'' together the
solutions for the pure phases in the presence of such singularities?


This question can be answered at the Landau-Ginzburg (LG) level, since
the latter affords one a simple expression for the local pressure,
which further simplifies in spherical geometry. The present treatment
of local pressure is based on a formal analogy between the
Landau-Ginzburg free energy functional and the mechanical action; the
spatial coordinate traversing the phase boundary in the former context
is analogous to the time in the latter context. In the case of
macroscopic coexistence, an LG-based treatment automatically yields
the familiar Pascal's law, which is formally analogous to energy
conservation in mechanics. For spherical droplets, on the other hand,
there is a pressure excess inside; this excess is well approximated by
the Laplace expression when the droplet is sufficiently large. We will
observe that the boundary conditions that must be imposed on the order
parameter at the singularity of the bulk free energy amount to a
continuity of local pressure and, in fact, hearken back to conditions
for macroscopic phase equilibrium elucidated by Gibbs, except that now
pressure may be explicitly coordinate-dependent.  The resulting
treatment allows one to explicitly build the free energy profile for
droplet nucleation as a function of the droplet radius and the
chemical composition at the phase boundary.
We establish that the composition at the droplet boundary changes
little during nucleation.

The developed formalism is next generalized to situations where the
local pressure is {\em not} spatially homogeneous even for flat
interfaces. Important examples of such situations are phase boundaries
pinned by an external potential; the situation can alternatively be
thought of as amphiphilic molecules collecting at phase boundaries. In
the resulting, simple treatment, we explicitly observe the effects of
the amphiphiles' activity on the renormalization of the surface
tension. We obtain explicit coordinate dependences of pressure and
chemical potential {\em within} the interface, as well as the explicit
form of the binding potential between the interface and the
detergent. The mathematical simplicity of the formalism allows one to
work both in equilibrium and away from equilibrium, in a steady-state
approximation for individual phases.

The above developments turn out to afford one an arguably simplified
perspective at yet another, deep problem that has been a subject of
much interest and controversy,~\cite{0034-4885-77-4-042501,
  biroli:12A301} namely the mechanism of activated transport in glassy
liquids.~\cite{LW_ARPC} We will observe that the down-renormalization
of the free energy barrier for activated transport in glassy liquids
can be thought of as a decrease in the surface tension of an
inter-phase boundary in the presence of uniformly seeded, randomly
oriented amphiphiles. The square-root scaling of the mismatch penalty
between dissimilar aperiodic structures with the size of the
reconfiguring region can be viewed as a consequence of the law of
large numbers. Likewise, the present formalism allows one to flesh out
the ``wetting'' view of the mismatch penalty, due to Xia and
Wolynes.~\cite{XW} Connections with other existing treatments of
glassy dynamics can be also made.

The article is organized as follows: In Section~\ref{review} we
provide a brief pedagogical review of how to set up the problem of
phase coexistence at the Landau-Ginzburg level. In
Section~\ref{pressure}, we derive an expression for local pressure for
the LG functional, which yields an explicit formula for spherical
geometries. There we establish Pascal's law and develop a systematic
treatment of local pressure that can be applied to mixtures and
off-equilibrium configurations. These results are used to obtain
explicit droplet solutions for liquid mixtures in Section~\ref{CNT}
and the problem of a flat interface exposed to a pinning potential
and/or amphiphilic adsorbents in Section~\ref{amphi}. These
developments provide two alternative perspectives on the problem of
renormalization of the mismatch penalty between dissimilar aperiodic
structures in glassy liquids, which is discussed in
Section~\ref{wettingSection}. The final Section~\ref{summary} briefly
summarizes the present findings.

\section{Phase Equilibrium and Phase Ordering at Landau-Ginzburg Level:
  Setup of the Problem}
\label{review}

Consider a Landau-Ginzburg free energy functional:
\begin{equation} \label{F} F(\{ \psi(\br) \}) = \int d^3 \br
  \left[\frac{\kappa}{2} (\nabla \psi)^2 + \cV(\psi) \right] \equiv \int
  d^3 \br f(\br),
\end{equation}
where $f(\br) \equiv \kappa (\nabla \psi)^2/2 + \cV(\psi)$ is thus the
local free energy density. The quantity $\cV(\psi)$ corresponds with
the bulk free energy density of a spatially uniform system $\psi =
\text{const} \Rightarrow \nabla \psi = 0$, since under these
circumstances $F/V = \cV$, where $V$ is the total volume. In contrast,
the {\em full} free energy density $f$ also accounts for the free
energy cost of spatial variations in the order parameter $\psi$. The
quantity $\psi$ is a suitable intensive variable of interest. The
continuum integral in Eq.~(\ref{F}) can be thought of as a limit of a
discrete sum over cells with fixed volumes, the cells tiling the
space. Thus $f$ must be regarded as a free energy computed at constant
volume, insofar as the order parameter reflects density
variations. Prominent examples of such types of free energy are given
by the Helmholtz free energy $A$:
\begin{equation} A = - k_B T \ln \sum_i e^{-E_i(V, N)/k_B T}
\end{equation}
and the grand canonical free energy $\Omega \equiv - pV$:
\begin{equation} \Omega \equiv - pV = - k_BT \ln \sum_i e^{-[E_i(V) -
    \mu N_i(V)]/k_B T}.
\end{equation}
In the Helmholtz case, the summation is over all possible microstates
$i$ such that $V_i = V$ and $N_i = N$, while in the grand-canonical
case, the particle number $N_i$ is unconstrained.~\cite{McQuarrie,
  LLstat} If, on the other hand, the degree of freedom of interest
$\psi$ is not strongly coupled to the density, as would be the case
for the magnetization in a magnet, the volume integration in
Eq.~(\ref{F}) largely amounts to a summation over sites hosting the
degree of freedom and does not put limitation on the specific choice
of free energy.  An interesting example of the latter situation is
represented by continuum mechanics~\cite{LLelast}, even though the
latter theory explicitly deals with deformations such as density
changes. In continuum mechanics, the integration is over the volume of
an {\em undeformed} sample and, thus, can be thought of as,
essentially, a sum over lattice sites, not integration over actual
space.

\begin{figure}[t]
  \includegraphics[width= \figurewidth]{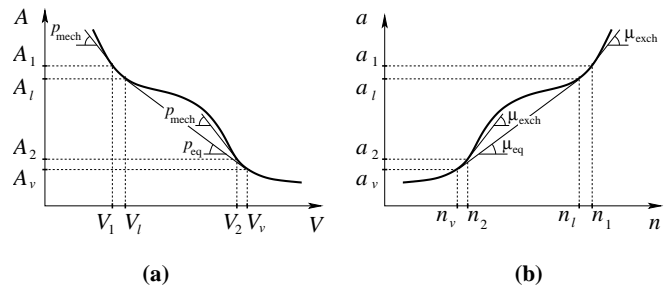}
  \caption{\label{bulk} {\bf (a)} Sketch of the Helmholtz free energy
    $A$ as a function of volume $V$ of a liquid at constant
    temperature and particle number, below the critical point. The
    double tangent corresponds to the free energy of the phase
    separated system and is followed in equilibrium. While the two
    states 1 and 2 are in mechanical equilibrium, they are not in
    equilibrium with regard to particle exchange, unless
    $p_\text{mech}$ is equal to the (negative) slope of the double
    tangent, whereby $V_1 = V_l$ and $V_2 = V_v$. {\bf (b)} Sketch of
    the Helmholtz free energy per unit volume, $a \equiv A/V$, as a
    function of particle concentration, at constant volume and
    temperature. States 1 and 2 are in equilibrium with respect to
    particle exchange, but not in mechanical equilibrium, unless
    $\mu_\text{exch}$ is equal to the slope of the double tangent.}
\end{figure}

When the order parameter corresponds to a conserved quantity---such as
the particle number---discontinuous phase transitions are signalled by
the presence of a sign change in the second derivative $\prtl^2
\cV/\prtl \psi^2$. Consider, for instance, the Helmholtz free energy
of a spatially homogeneous van der Waals gas below the critical point,
at fixed temperature and particle number. The free energy can be
computed by formal integration $A_2 = A_1 - \int_1^2 p dV$ of the van
der Waals isotherm $p = N k_B T/(V - Nb) - a(N/V)^2$, see sketch in
Fig.~\ref{bulk}(a).  At any value of the volume $V$ intermediate
between the boundaries $V_l$ and $V_v$ of the non-concave region in
the $A(V)$ dependence, the system can lower its Helmholtz free energy
by phase separating into vapor (mole fraction $x_v = (V - V_l)/(V_v -
V_l)$) and liquid (mole fraction $x_l = (V_v - V)/(V_v - V_l) = 1 -
x_v$). Upon phase separation, the dependence of the free energy on
volume is given by the straight line $A = x_v A_v + x_l A_l$, which is
the double tangent line that runs {\em under} the $A(V)$ curve that
was calculated under the assumption of complete spatial uniformity of
the gas. We remind the reader that the Helmholtz free energy $A = E -
TS$ must be at its minimum at constant temperature and
volume.~\cite{LLstat} Viewed this way, the entropy $S = - k_B \sum_i
p_i \ln p_i$ ($\sum_i p_i = 1$) is understood in its Gibbsian sense,
where the probabilities $p_i$ represent the statistical weights of
microstates $i$ and can be computed under a constraint of one's
liking.  The curved and flat $A(V)$ segments at $V_l < V < V_v$ in
Fig.~\ref{bulk}(a) are computed, respectively, with and without
imposition of the constraint that the system be spatially homogeneous.

An infinite number of configurations 1 and 2 such that $-\prtl A/\prtl
V|_1 = -\prtl A/\prtl V|_2 \Rightarrow p_1 = p_2 \equiv p_\text{mech}$
correspond to a mechanical equilibrium, see the short, parallel
tangents in Fig.~\ref{bulk}(a). Yet only one of those configurations,
viz., $(A_2 - A_1)/(V_2 - V_1) = \prtl A/\prtl V|_1 = \prtl A/\prtl
V|_2$ also corresponds to {\em full} equilibrium, i.e., when the two
parallel tangents strictly coincide thus implying $V_1 = V_l$ and $V_2
= V_v$. Indeed, only at the corresponding pressure $p_\seq \equiv -
(A_v - A_l)/(V_v - V_l)$ do the Gibbs free energies $G = A + pV$ and,
hence, the bulk chemical potentials of the two phases become equal:
$A_1 + p_1 V_1 = A_2 + p_2 V_2 \Rightarrow \mu_1 N = \mu_2 N$, since
$G = \mu N$.~\cite{LLelast}

In the context of the density functional in Eq.~(\ref{F}), it is
practical to vary the particle number at constant volume; density
fluctuations are thus viewed as fluctuations of the particle number at
fixed volume, not fluctuations of the local specific volume. The
sketch in Fig.~\ref{bulk}(b) of the Helmholtz free energy density $a
\equiv A/V$ as a function of concentration $n = N/V$, at constant
volume, demonstrates that there are an infinite number of ways for two
phases to co-exist in which there is no particle exchange between the
phases: $\prtl a/\prtl n|_1 = \prtl a/\prtl n|_2 \Rightarrow \mu_1 =
\mu_2 \equiv \mu_\text{exch}$. Yet only one of these configurations is
also in {\em mechanical} equilibrium, so that $(a_2 - a_1)/(n_2 - n_1)
= \prtl a/\prtl n|_1 = \prtl a/\prtl n|_2$. Hereby, $A_1 - \mu_1 N_1 =
A_2 - \mu_2 N_2 \Rightarrow - p_1 V = -p_2 V$. At full equilibrium,
$\mu_1 = \mu_2 = (a_l - a_v)/(n_l - n_v) = \mu_\seq$.

\begin{figure}[t]
  \includegraphics[width= 0.65 \figurewidth]{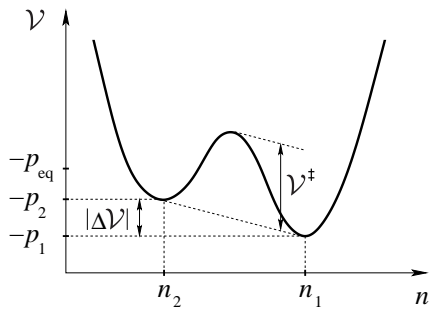}
  \caption{\label{V} Sketch of the bulk free energy density $\cV$ from
    Eq.~(\ref{F}), as could be obtained, for instance, from the
    Helmholtz free energy density by making the equilibrium state also
    the standard state. In the latter case, one sets $\cV = a - \mu_0
    n$, where $a$ is the thick curve from Fig.~\ref{bulk}(b), and
    $\mu_0 = \mu_\text{exch}$.}
\end{figure}

The value of the chemical potentials $\mu_1 = \mu_2$ in
Fig.~\ref{bulk}(b) has been chosen, for concreteness, so that the bulk
vapor is oversaturated with respect to the liquid. The notion of the
metastability of the vapor state becomes particularly vivid if one
stipulates that either state 1 or 2 be the reference---or {\em
  standard}---state with regard to the chemical potential. (The choice
of the standard state is, of course, arbitrary and is made according
to one's convenience.)  Formally, making state 1 the standard state is
accomplished by using the free energy density $\cV \equiv a - \mu_0 n$
where the fixed quantity $\mu_0$ is numerically equal to the chemical
potential at concentration $n_1$ (or $n_2$) according to the original
Helmholtz free energy $\mu_0 = \prtl a/\prtl n|_{n_1} = \prtl a/\prtl
n|_{n_2}$.  The equilibrium values of the concentration---stable or
metastable---thus correspond to the minima of the function $\cV$ by
construction: $\prtl \cV/\prtl n|_1 = \prtl \cV/\prtl n|_2 = 0$.  The
free energy density $\cV$ corresponding to Fig.~\ref{bulk}(b) is
sketched in Fig.~\ref{V}.  One can alternatively think of $\mu_0$ as
an externally imposed chemical potential. Hereby, the actual (bulk)
chemical potential $\mu$ is equal to $\mu_0$ only in equilibrium,
i.e., at the minima of $\cV$. Because at the minima, $\cV(n) = (A -
\mu_0 N)/V = (A - \mu N)/V = -p$, the depths of the minima of $\cV$
are numerically equal to the negative pressure. Note that when the
vapor and liquid are in equilibrium with respect to particle exchange
but not in mechanical equilibrium, as in Fig.~\ref{V}, the pressure is
always greater in the undersaturated phase, as expected, while the
magnitude of the deviation $|p - p_\seq|$ from the equilibrium value
$p_\seq$ is always greater in the denser phase.

The free energy density in Fig.~\ref{V} is a typical example of the
free energy density $\cV$ from Eq.~(\ref{F}) in the context of a {\em
  discontinuous} phase transition. Hereby, the order parameter has
distinct values in the distinct phases of interest; the two phases
correspond to the two minima of the function $\cV$, which are
separated by a barrier. Another common example of a conserved order
parameter is magnetization, call it $\cal M$. In this case, the role
of the chemical potential $\mu$ is played by the externally imposed
magnetic field $h$, since $dA = - S dT + h d \cal M$. The transition
itself amounts to a reversal of the magnetization of a magnet cooled
below its Curie point. In the rest of the paper, we largely limit
ourselves to fluid mixtures made of $M$ components whose
concentrations we denote with $n_i$, $i = 1 \ldots M$.

We specifically allow the coefficient $\kappa$ at the square gradient
term in Eq.~(\ref{F}) to explicitly depend on the order
parameters. (The dependence of $\kappa$ on the coordinate is thus
exclusively through the coordinate dependence of the order parameter,
if any.)  This coefficient directly reflects particle-particle
interactions and thus generally differs between distinct phases of
matter. Specifically, the coefficient $\kappa$ (for a pure substance)
scales linearly with the direct correlation function $c(\br_1, \br_2)
\equiv -\beta \delta^2 F_\text{ex} [n(\br)]/\delta n (\br_1) \, \delta
n (\br_2)$, where the derivative is evaluated at equilibrium density
and $F_\text{ex}$ is the non-ideal portion of the free energy.  The
direct correlation function in a fluid generally scales with the bulk
modulus and thus sensitively depends on the density and temperature.
Explicit calculations for the coefficient $\kappa$ can be found in
Refs.~\cite{RL_sigma0, L_AP}

We thus adopt the following Landau-Ginzburg functional:
\begin{equation} \label{Fmulti} F(\{ n_i(\br) \}) = \int d^3 \br
  \left[\sum_{ij}^M \frac{\kappa_{ij}}{2} (\bm \nabla n_i \bm \nabla
    n_j) + \cV(\{ n_i \}) \right]
\end{equation}
where the summation is over distinct chemical species. The matrix
$\kappa_{ij}$ is automatically symmetric, $\kappa_{ij} = \kappa_{ji}$,
and, also, positive-definite by construction.

The free energy of the configuration $n_i(\br) + \delta n_i(\br)$
within a connected region, relative to that of configuration
$n_i(\br)$, is given to the first order in a (small) quantity $\delta
n_i$ by the expression:
\begin{align} \label{dFn} &\delta F \equiv F(\{ n_i + \delta n_i \}) -
  F(\{ n_i \}) \nonumber \\ &= \sum_{ij} \int_S (d \bm S \, \bm \nabla
  n_j) \kappa_{ij} \delta n_i + \sum_i \int_V dV \mu_i \delta n_i,
\end{align}
where the first and second integrals are over the boundary and the
bulk of the region, respectively, and
\begin{equation} \label{mu} \mu_i(\br) = \frac{\prtl \cV}{\prtl n_i} -
  \sum_j \bm \nabla (\kappa_{ij} \bm \nabla n_j) + \frac{1}{2}
  \sum_{lm} \frac{\prtl \kappa_{lm}}{\prtl n_i} (\bm \nabla n_l \bm
  \nabla n_m).
\end{equation}

By Eq.~(\ref{dFn}), the quantity
\begin{equation} \label{mui} \mu_i(\br) = \frac{ \delta F}{\delta n_i
    (\br) }.
\end{equation}
It thus equals the free energy cost of adding a particle to the system
at point $\br$, i.e., the local chemical potential. We see from
Eq.~(\ref{mu}) that in addition to the obvious contribution $\prtl
\cV/\prtl n_i$, the local chemical potential also exhibits
contributions due to the free energy cost of density variations.

If the local cost of adding a particle is not spatially homogeneous,
particle fluxes will emerge spontaneously. In spatial dimensions three
and higher, phenomenological Fick's law can be used to connect the
particle fluxes $\bm j_i$ and the gradients of the chemical potentials
of the species present:
\begin{equation} \label{Fick} \bm j_i = - \sum_j \widetilde D_{ij} \bm
  \nabla \mu_j,
\end{equation}
where the symmetric~\cite{PhysRev.37.405, PhysRev.38.2265,
  Langer19751225} matrix $\widetilde D_{ij}$ represents a set of
self-diffusion coefficents.  These quantities are generally
concentration-dependent and can be related to the regular
diffusivities through compressibilities. For instance,
Eq.~(\ref{Fick}) yields for a pure substance near equilibrium: $\bm j
= - \widetilde D (\prtl \mu/\prtl n) \bm \nabla n$, while $(\prtl
\mu/\prtl n) = v^2/\kappa_T$, where $\kappa_T$ and $v$ are the
compressibility and specific volume respectively. Consequently, the
ordinary diffusivity $D_\text{diff} = \widetilde D v^2/\kappa_T$.

Eq.~(\ref{Fick}), together with particle conservation $\dot n_i = -
\bm \nabla \bm j_i$, allows one to write down a system of equations
governing relaxation of the system toward equilibrium~\cite{Bray}:
\begin{equation} \label{nidot} \dot n_i = \sum_j \bm \nabla \left(
    \widetilde D_{ij} \bm \nabla \mu_j \right),
\end{equation}
with $\mu_i$ from Eq.~(\ref{mu}).  

By Eq.~(\ref{Fick}), the chemical potential must be spatially uniform
in full equilibrium:
\begin{equation} \label{mu0} \mu_i = \text{const} = 0 \text{, in
    equilibrium,}
\end{equation}
where we have used that steady state is also the standard state, by
construction. However, the condition for {\em local} steady state is
much less restrictive. According to Eq.~(\ref{nidot}), one must only
require that $\sum_ j \bm \nabla ( \widetilde D_{ij} \bm \nabla \mu_j
) = 0$, which allows for spatially varying chemical potentials.

To avoid confusion, we note that Eq.~(\ref{nidot}) is incomplete in
two ways, in addition to being phenomenological. For one thing, it
does not account for advection, which is generally present in
mixtures.~\cite{Bray} Indeed, an individual component can flow also
with the mixture as a whole, not only by itself.  Still, this
complication does not arise in steady state, when there is no net
flow. Also, Eqs.~(\ref{nidot}) should properly contain terms
accounting for thermal noise~\cite{ELS1996} to describe processes
other than relaxation toward equilibrium.


Although less common, non-conserved order parameters are encountered
in applications too. In the absence of a conservation law for the
order parameter, one often uses the empirical Landau-Onsager
ansatz~\cite{LLstat, PhysRev.37.405, PhysRev.38.2265, Bray,
  Goldenfeld} for the time evolution of the
system:~\cite{LandauKhalatnikov1954}
\begin{equation} \label{LO} \dot{\psi} = - {\cal K} \frac{\delta
    F}{\delta \psi}.
\end{equation}
Eq.~(\ref{LO}) is the simplest empirical law one can write down that
relates the rate of relaxation toward or away from steady state to the
deviation $\delta \psi(\br) \equiv \psi(\br) - \psi_0(\br)$ of the
order parameter $\psi(\br)$ from its steady-state value $\psi_0(\br)$
that optimizes the functional: $ \delta F/\delta \psi|_{\psi_0} = 0$.
Finally note that Eqs.~(\ref{mui}) and (\ref{mu0}) on the one hand and
Eq.~(\ref{LO}) on the other hand imply the equations determining the
steady state themselves are identical for conserved and non-conserved
order parameters, even though relaxation toward or away from steady
state is governed by distinct laws in the two cases.

%

\section{Pressure in the Landau-Ginzburg Functional}
\label{pressure}

Stationary points of the functional that are also {\em minima}
correspond to a spatially uniform system residing wholly in a specific
phase. Such a free energy minimum will be either stable or metastable
according to whether the corresponding minimum in the bulk free energy
density $\cV(\psi)$ is stable or metastable. In the following, we will
be discussing additional stationary points of the functional that
correspond not to minima but to {\em saddle points}. Such saddle
points, if any, are transition-state configurations that could arise
during a phase transition between two phases. Specifically, we
consider {\em droplet} configurations such that the order parameters
achieve their bulk values $n_i^{(\text{out},b)}$ in the majority
phase, at infinity. We demand that near the origin, the mixture is on
the other side of the transition state of the bulk free energy density
$\cV$, i.e., in the minority phase.

In the limit of thin interface, the transition-state droplet
configuration is obviously spherically symmetric, since this
configuration minimizes the surface area for a given volume of the
droplet. Alternatively said, such spherically-symmetric configuration
satisfies Pascal's law within individual phases because it insures
uniformity of pressure (separately) on the inside and outside.  Below
we assume that the bulk free energy density does not explicitly depend
on the coordinate, $\prtl \cV/\prtl \bm r = 0$, unless explicitly
stated otherwise. Now, when the interface is not thin, it appears
difficult to make general statements about the symmetry of the droplet
solution because Eqs.~(\ref{mu0}) are non-linear. Still, inasmuch as
we expect the lowest free energy saddle point to be unique we can also
expect the corresponding solution to be spherically symmetric. Indeed,
suppose on the contrary that the solution is only of {\em discrete}
symmetry with respect to rotations. Further, suppose for the sake of
argument that this solution transforms under the corresponding
symmetry operations in the same way as the coordinate $x$. For each
such solution, there will be also two other solutions, transforming as
the coordinates $y$ and $z$ respectively, which contradicts the
original assumption that the solution is unique. Cylindrically---but
not spherically---symmetric solutions can be dismissed by the same
token. The following boundary conditions thus apply to spherically
symmetric droplets:
\begin{equation} \label{bc} \begin{array}{rcl} n_i (\br = \infty) &=&
    n_i^{(\text{out},b)}
    \\     \\
    \bm \nabla n_i (\br = 0) &=& 0.
    \end{array}  
\end{equation}
The actual values of the order parameters may not reach their bulk
values $n_i^{(\text{in},b)}$ in the minority phase. However in view of
the reflection symmetry of the problem, their spatial derivatives must
vanish at the origin to avoid a discontinuity.


In the presence of spherical symmetry, Eq.~(\ref{mu}) yields in $D$
spatial dimensions:
\begin{align} \label{ELsph} &\sum_{j} \kappa_{ij} \left( \frac{d^2
      n_j}{d r^2} + \frac{D-1}{r} \frac{d n_j}{d r} \right) \\ &=
  \frac{\prtl \cV}{\prtl n_i} - \sum_j \frac{d \kappa_{ij}}{d r} \,
  \frac{d n_j}{d r} + \frac{1}{2} \sum_{lm} \frac{\prtl
    \kappa_{lm}}{\prtl n_i} \, \frac{d n_l}{d r} \, \frac{d n_m}{d r}
  - \mu_i. \nonumber
\end{align}
Note the dependence of $\kappa_{ij}$ on the coordinate is exclusively
through the coordinate dependences of the concentrations.

Multiplying Eq.~(\ref{ELsph}) by $dn_i$, summing over $i$, and
integrating from some point 1 to point 2 yields:
\begin{align} \label{12} \sum_{ij} & \left. \frac{\kappa_{ij}}{2}
    \frac{d n_i}{d r} \frac{d n_j}{d r} \right|_1^2 - \cV |_1^2 \\ &=
  - (D-1)\int_1^2 \frac{dr}{r} \sum_{ij} \kappa_{ij} \frac{d n_i}{d r}
  \frac{d n_j}{d r} - \sum_i \int_1^2 \mu_i \, dn_i.  \nonumber
\end{align}
In deriving the above equation, we have taken advantage of the
symmetry of the $\kappa_{ij}$ matrix. One immediately observes that
when $D=1$, i.e., for a flat interface, and in complete equilibrium,
$\mu_i = 0$, the following quantity is conserved:
\begin{equation} \label{cons} \sum_{ij} \frac{\kappa_{ij}}{2} \frac{d
    n_i}{d r} \frac{d n_j}{d r} - \cV = \text{const, } \hspace{3mm}
  D=1, \, \mu_i = 0.
\end{equation}
This is expected based on the formal correspondence between the free
energy function (\ref{Fmulti}) in the $D=1$, $\mu_i = 0$ case and the
action in classical mechanics,~\cite{LLmech} i.e. $\int dt
[\Sigma_{ij} M_{ij} \dot x_i \dot x_j/2 - U(x_i)]$. Hereby, the
concentrations $n_i$ are the analogs of the mechanical coordinates
$x_i$, the coordinate $x$ of time $t$, and the bulk free energy
density $\cV$ of the negative potential energy $-U$. The conserved
quantity (\ref{cons}) is thus formally analogous to energy in
mechanics.

Taking the above, formal correspondence with mechanics a step further,
we note that the mechanical energy is the partial (not full!)
derivative of the mechanical action with respect to time, with the
minus sign.~\cite{LLmech} We can use this notion to establish that the
analog of the mechanical energy in the present context is actually the
{\em mechanical pressure}. The latter statement turns out to be
correct not only in 1D, but, more generally, in any number of
dimensions so long as the steady state solution is spherically
symmetric. Indeed, consider first an arbitrary droplet geometry. It
will be convenient to switch to a modified free energy function
$\widetilde F$:
\begin{equation} \label{Ftilde} \widetilde F (\{ n_i \}) \equiv F -
  \sum_i \int \mu_i \, n_i \, dV = \int (f - \sum_i \mu_i \, n_i) dV,
\end{equation}
where $\mu_i$'s are now regarded as some preset functions of the
coordinate $\br$. The quantity $f$ is the Helmholtz free energy
density corresponding to the functional in Eq.~(\ref{Fmulti}). The
free energy of the configuration $n_i(\br) + \delta n_i(\br)$ within a
connected region, relative to that of configuration $n_i(\br)$, is
given to the first order in a (small) quantity $\delta n_i$ by the
expression:
\begin{align} \label{dFmu} &\delta \widetilde F \equiv \widetilde F(\{
  n_i + \delta n_i \}) - \widetilde F(\{ n_i \}) \\ &= \sum_{ij}
  \int_S (d \bm S \, \bm \nabla n_j) \kappa_{ij} \delta n_i + \sum_i
  \int_V dV \left(\frac{\delta F}{\delta n_i} - \mu_i \right) \delta
  n_i. \nonumber
\end{align}
We now limit our attention to those specific concentration profiles
that satisfy Eq.~(\ref{mui}).  For those specific profiles, the
function $\widetilde F$ is actually a Legendre transform of the
functional $F$, namely, the grand-canonical free energy as defined for
a coordinate-dependent chemical potential:
\begin{equation} \Omega(\{ \mu_i \}) \equiv F - \sum_i \int \mu_i \,
  n_i \, dV
\end{equation}
Under these circumstances, the volume integral vanishes in
Eq.~(\ref{dFmu}); one may present the free energy variation as an
integral over the sample's boundary.

Let us now choose as our boundary an {\em isobaric} surface, or
surface of constant pressure. This way, the hydrostatic forces near
the boundary are exactly normal to the latter. Varying the free energy
with respect to the displacement $l$ along the normal to the surface
at a specific location thus simply yields the negative pressure at
that location times the area $S$ of the (small) patch over which the
variation is performed:
\begin{equation}
  \frac{1}{S} \left( \frac{\prtl \Omega}{\prtl l} \right)_{T \! , \, \mu_i} 
  = \left( \frac{ \prtl \Omega}{\prtl V} \right)_{T \! , \, \mu_i}  = -p.  
\end{equation}
On the other hand, by the chain rule of differentiation:
\begin{equation} \frac{d \widetilde F}{d l} = \frac{\prtl \widetilde
    F}{\prtl l} + \sum_i \int_S d S(\br) \, \frac{\delta \widetilde
    F}{\delta n_i(\br)} \, \frac{d n_i(\br)}{d l},
\end{equation}
where $\delta \widetilde F/\delta n_i(\br)$ is the functional
derivative of the free energy with respect to the concentration
$n_i(\br)$ at point $\br$ belonging to the boundary. According to
Eqs.~(\ref{dFmu}) and (\ref{mui}), this derivative is given by the
expression $\sum_j \kappa_{ij} (\bm e(\br) \, \bm \nabla n_j)$, where
$\bm e(\br) \equiv d \bm S(\br)/d S(\br)$ is the external normal to
the boundary at the point $\br$, $e^2 =1$. Note that $dn_i/dl = (\bm
e(\br) \, \bm \nabla n_i)$. Finally, the full derivative $d \widetilde
F/dl$ is simply given by the integrand in Eq.~(\ref{Ftilde})
integrated over the patch, according to the Newton-Leibniz
formula. Putting these notions together, while taking the limit $S \to
0$, yields the following formula for the pressure in the
Landau-Ginzburg functional:
\begin{equation} \label{p} p = - \cV + \sum_i \mu_i n_i + \sum_{ij}^M
  \kappa_{ij} \left[(\bm e \bm \nabla n_i) (\bm e \bm \nabla n_j) -
    \frac{1}{2}(\bm \nabla n_i \bm \nabla n_j) \right],
\end{equation}
where $\bm e \equiv d \bm S/d S$, as before. Note the above equation
is invariant with respect to the choice of the standard state for the
chemical potential, as it should be.

Eq.~(\ref{p}), by itself, cannot be used to determine the pressure in
arbitrary geometries, unless the orientation of the isobaric surface
happens to be known. Some insight may be already gleaned, however,
when the $\kappa_{ij}$ matrix is diagonal, in which case it is easy to
write an upper bound for the pressure that does not depend on that
orientation:
\begin{equation} \label{pmax} p \le - \cV + \sum_i \mu_i n_i +
  \sum_{ij}^M \frac{\kappa_{ij}}{2}(\bm \nabla n_i \bm \nabla n_j) ,
  \text{ if } \kappa_{ij} = \kappa_i \delta_{ij}.
\end{equation}
Note that Eqs.~(\ref{dFmu})-(\ref{pmax}) also apply when the bulk free
energy density $\cV$ explicitly depends on the coordinate.

Fortunately, an explicit expression for pressure can be obtained in
the presence of spherical symmetry, whereby the isobaric surfaces also
coincide with the surfaces of constant concentrations and chemical
potentials. Consequently, the gradients $\bm \nabla n_i$ are strictly
parallel to the external normal $\bm e$ and so Eq.~(\ref{p}) yields:
\begin{equation} \label{psph} p(r) = - \cV + \sum_i \mu_i n_i +
  \sum_{ij} \frac{\kappa_{ij}}{2} \frac{d n_i}{d r} \frac{d n_j}{d r},
\end{equation}
c.f. Eq.~(\ref{pmax}).  We thus observe that the quantity
(\ref{cons}), which is conserved for flat interfaces (when $\mu_i =
0$), corresponds with the mechanical pressure. The spatial homogeneity
of pressure for a flat interface---and, hence, during {\em
  macroscopic} phase coexistence---is, of course, the familiar
Pascal's law.

Thus, Eqs.~(\ref{12}) and (\ref{psph}) yield for an equilibrium flat
interface perpendicular to the $x$ axis:
\begin{equation} \label{pflat} p = -\cV + \sum_{ij}
  \frac{\kappa_{ij}}{2} \frac{d n_i}{d x} \frac{d n_j}{d x} =
  \text{const}, \hspace{5mm} D=1, \, \mu_i = 0.
\end{equation}
One may think of this equation as saying that the variation in the
order parameter exactly compensates for the decrease in pressure, due
to the bulk free energy term $\cV$, that arises in the transition
state.
This is not unlike how the variation in the chemical potential, due to
the bulk free energy term, is exactly compensated by the spatial
variation in the order parameter in Eqs.~(\ref{mu}) and (\ref{mu0}).
No such pressure compensation takes place in spatial dimensions
greater than one, however, for which Eqs.~(\ref{12}) and (\ref{psph})
yield:
\begin{align} \label{vdW} p_2 - p_1 = - (D-1)\int_1^2 \frac{dr}{r}
  \sum_{ij} \kappa_{ij} \frac{d n_i}{d r} \frac{d n_j}{d r} + \sum_i
  \int_1^2 \!\!\!\! n_i \, d \mu_i. 
\end{align}
This is a generalization of a result first obtained by van der
Waals~\cite{vdWinterf} for a single-component substance and a constant
$\kappa$ to mixtures, concentration-dependent coefficients
$\kappa_{ij}$, and off-equilibrium situations in the sense that $\mu_i
\ne 0$. Another way to look at formula (\ref{vdW}) is to recall that
in an equilibrated, uniform sample, $G = \sum_i \mu_i N_i$ while $dG =
-S dT + V dp + \sum_i \mu_i d N_i$. At constant temperature and
volume, this yields $p_2 - p_1 = \sum_i \int_1^2 n_i \, d \mu_i$ for
any pair of equilibrium states 1 and 2. According to Eq.~(\ref{vdW}),
such two states can be representative of the very same sample, however
the expression for the pressure difference must be corrected for a
finite curvature of density variation patterns, if any.

Now choose points 1 and 2 at the center of the droplet and infinity,
respectively. In view of positive-definiteness of the matrix
$\kappa_{ij}$, Eq.~(\ref{vdW}) shows that pressure always increases
monotonically toward the center of a spherical droplet in full
equilibrium. (We remind the reader that the equilibrium may be
unstable, as is the case with critical nuclei.)  Furthermore, in the
limit of infinitely weak undersaturation of the minority phase,
$\Delta \cV \to 0$, the expression for the excess pressure inside the
droplet, relative to the bulk, reduces to the venerable Laplace
form. Indeed, sufficiently close to macroscopic coexistence of the two
phases, the droplet size can be made arbitrarily larger than the
interface width. The latter width must be finite and, in fact, is
limited by its value during macroscopic co-existence: According to
Eq.~(\ref{12}), the derivatives $d n_i/d r$ become only larger when $D
> 1$. On the other hand, the interface width during macroscopic
coexistence is finite because otherwise, the system could always lower
its free energy by an infinite amount, by locally ``falling'' into one
of the two minima in Fig.~\ref{V}. Under these circumstances, the
order parameters differ from their (spatially-uniform) values in the
bulk phases in a spatial interval that can be made arbitrarily
narrower than the droplet size. Consequently one can unambiguously
define a droplet radius $R$. Because the derivatives $d n_i/dr$ have
appreciable magnitudes only within a finite interval, the $1/r$ factor
in the integrand of Eq.~(\ref{12}) can be taken outside the integral,
while the integral itself can be replaced by the value it achieves in
macroscopic equilibrium, $\Delta \cV = 0$, i.e., for a flat
interface. Further, Eq.~(\ref{cons}) implies that the integral is in
fact equal to the excess free energy of the flat interface, relative
to the spatially uniform state, per unit area. This integral is thus
equal to the surface tension coefficient $\sigma$:
\begin{equation} \label{sigma} \sigma = \int_{-\infty}^\infty dx
  \sum_{ij} \kappa_{ij}\frac{d n_i}{d x} \frac{d n_j}{d x} =
  \int_{-\infty}^\infty dx \, (\cV - \cV_\infty),
\end{equation}
where, we remind, $\Delta \cV = 0$. Thus according to this argument
and Eq.~(\ref{vdW}), for any two points 1 and 2 inside and outside the
droplet respectively, we obtain the venerable Laplace's formula $p_1 =
p_2 + (D-1) \sigma/R$. This result, combined with Fig.~\ref{bulk}(b)
yields two classic notions: On the one hand, the minority phase must
be undersaturated relative to the majority phase. On the other hand,
the result allows one to estimate the deviation of pressure in either
phase compared to its value during macroscopic coexistence, which then
yields the venerable Gibbs-Thompson formula.~\cite{LLstat}

For a single-component fluid, Eq.~(\ref{sigma}) can be significantly
simplified:~\cite{CahnHilliard}
\begin{equation} \label{sigma1} \sigma = \int_{n_{\mathrm{min},
      1}}^{n_{\mathrm{min},
      2}}  dn \sqrt{2 \kappa (\cV-\cV_\infty)},
\end{equation}
where $n_{\mathrm{min}, 1}$ and $n_{\mathrm{min}, 2}$ correspond to
the two minima of the bulk free energy $\cV(n)$; $n_{\mathrm{min}, 1}
< n_{\mathrm{min}, 2}$ by construction.  The above expression is
remarkable in that it allows one to compute the surface tension
coefficient (for a flat interface) without the need to solve
Eq.~(\ref{mu0}). By Eq.~(\ref{sigma1}), the surface tension
coefficient is determined, generically, by the height $\cV^\ddagger$
of the barrier in the bulk free energy term, see Fig.~\ref{V}, and the
value of the coefficient $\kappa$ around the transition state: $\sigma
\simeq \sqrt{\kappa^\ddagger \cV^\ddagger}$. The corresponding length
scale, which reflects the interface width, is $l \simeq
\sqrt{\kappa^\ddagger/\cV^\ddagger}$, and the surface tension
coefficient is simply $\sigma \simeq \cV^\ddagger l$. Some
contributions to both the interface width and the surface tension
coefficient are expected from the order parameter fluctuations near
the minima. The corresponding lengthscale is proportional to
$[\kappa/(\prtl^2 \cV/\prtl n^2)]^{1/2}$. These contributions,
however, must not be very significant, otherwise the system must be
regarded as being close to criticality, in which case the mean-field
Landau-Ginzburg functional becomes inadequate either quantitatively
or, sometimes, even qualitatively.

In a mixture, the surface tension coefficient can still be presented
as a line integral connecting the free energy minima that correspond
with the phases in question, in the $\{n_i \}$ space. This is obvious
from the l.h.s. equation of Eq.~(\ref{sigma}). While the resulting
expression allows one to decompose the overall mismatch free energy
into contributions from individual components, it requires one to
first solve Eq.~(\ref{mu0}), which is generally not a simple task.

\section{Singular Bulk Free Energy Density.  Steady State Solution.}
\label{CNT}

\begin{figure}[t]
  \includegraphics[width=  \figurewidth]{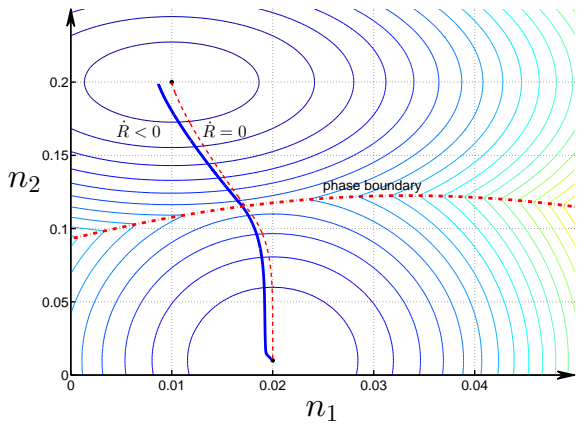}
  \caption{\label{n1n2V} Contour plot of the bulk free energy density
    $\cV(n_1, n_2)$ as a function of the concentrations $n_1$ and
    $n_2$ of the components of a binary mixture. The upper-left
    minimum corresponds with the undersaturated, minority phase. The
    two ``trajectories'' correspond to the density profiles of the two
    droplet solutions that are illustrated in detail in
    Fig.~\ref{munpr}. The parameters employed in this Section are as
    follows: $\kappa_{11}^\tin = 2000$, $\kappa_{22}^\tin = 300$, $
    \kappa_{11}^\tout = 20$, $\kappa_{22}^\tout = 15$, $\kappa_{12} =
    0$, $m_{11}^\tin = 400$, $m_{11}^\tout = 700$, $m_{22}^\tin = 40$,
    $m_{22}^\tout = 20$, $m_{12} = 0$, $n_1^{(\tin, b)} = 0.01$,
    $n_2^{(\tin, b)} = 0.2$, $n_1^{(\tout, b)} = 0.02$, $n_2^{(\tout,
      b)} = 0.01$, $\widetilde{D}_{12} = 0$, $\widetilde{D}_{ii}
    m_{ii} = 1$ for $i = 1, 2$ in both phases, $\Delta \cV =
    -0.04$. The units are arbitrary; as a rough guide, the unit of
    length is set at molecular dimensions and the unit of energy at a
    small fraction of $k_B T$. }
\end{figure}

Equations (\ref{mu0}) are non-linear because generally, the bulk free
energy $\cV$ is not a quadratic or linear function of the
concentrations while the coefficients $\kappa_{ij}$ may have distinct
values between the phases and interfacial regions. Although the
equations can be solved numerically, avoiding numerical instabilities
requires extra effort.  Since the actual functional form of $\cV$ and
$\kappa_{ij}$'s are usually not known in the first place, a practical
strategy is to make $\cV$ a quadratic function of the concentrations
$n_i$ everywhere in the $\{ n_i \}$ space except in a subspace of
lower dimensions. And so for each phase $\alpha$:
\begin{equation} \label{Vpara} \cV^{(\alpha)} = \sum_{ij}
  \frac{m_{ij}^\alpha}{2}(n_i - n_i^{(\alpha, b)})(n_j - n_j^{(\alpha,
    b)}) + \cV^{(\alpha, b)},
\end{equation}
where $n_i^{(\alpha, b)}$ denotes the bulk value of the concentration
of component $i$ in phase $\alpha$ and the symmetric matrix $m_{ij}$
is positive definite. The quantity $\Delta \cV \equiv
\cV^{(\text{in},b)} - \cV^{(\text{out},b)}$ thus gives the bulk free
energy difference, per unit volume, between the two phases, see
Fig.~\ref{V}.  For two coexisting phases, the bulk free energy $\cV$
consists of two paraboloids. If one chooses the phase boundary (in the
$\{ n_i \}$ space) as the intersection of the paraboloids, see
illustration in Fig.~\ref{n1n2V}, the resulting bulk free energy
experiences a discontinuity in the gradient. Otherwise, the
discontinuity is in the function itself.  The functional form
(\ref{Vpara}) presents some limitations in terms of modelling because
it does not allow one to vary the curvatures of the bulk free energy
surface in pure phases independently from the height of the barrier
separating the minima corresponding to the pure phases. This
effectively implies that the lengthscales $\sqrt{\kappa/m}$ and
$\sqrt{\kappa^\ddagger/\cV^\ddagger}$, discussed above, are not
independent.  Yet we shall see that even this simplified form of $\cV$
captures the essential features of thick interfaces that are not
accessible to the thin-wall approximation of the classical nucleation
theory.

A trajectory in the $\{ n_i \}$ space corresponding to droplet
configurations crosses the phase boundary in that space exactly once;
trajectories with more crossings automatically correspond to higher
free energies. As a result, the {\em physical} space is divided into
two parts separated by a sharp spherical boundary at some radius
$r=R$, the inside and outside parts corresponding to the minority and
majority phase respectively.  Likewise, one may regard $\kappa_{ij}$'s
as constant in each phase but assume they may experience a
discontinuous jump at the phase boundary. Under these circumstances,
Eqs.~(\ref{mu0}) become linear within individual phases:
\begin{equation} \label{mu0linear} \sum_j \kappa_{ij} \nabla^2 n_j =
  \sum_j m_{ij} (n_j - n_j^{(b)})
\end{equation}
while the following boundary conditions must be specified at the
dividing surface:
\begin{align}
  &n_i(R^+) = n_i(R^-) \equiv n_i^\ddag  \label{eq:BCconc}\\
  & \left. \left( \sum_{ij} \frac{\kappa_{ij}}{2} \frac{d n_i}{d r}
      \frac{d n_j}{d r} - \cV \right) \right|_{R^-}^{R^+} =
  0 \label{eq:BCp}
\end{align}
The first set of conditions, Eq.~(\ref{eq:BCconc}), ensures continuity
of the concentrations at the dividing surface. These conditions are
chosen based on the physical consideration that in a fluid, molecules
must be able to exchange positions and so a fluid can be only defined
within a finite region, even if very small. Consequently, the
concentration must change {\em continuously} in space. To give a
counterexample, the effective interface between a {\em solid} and the
corresponding vapor can be made arbitrarily narrow sufficiently below
the triple point.

The second condition, Eq.~(\ref{eq:BCp}), which places a constraint on
the concentration gradients at the dividing surface, stems from
Eq.~(\ref{12}). According to Eq.~(\ref{psph}), the condition in
Eq.~(\ref{eq:BCp}) corresponds with the continuity of pressure.  This
is required so that there is a balance of microscopic forces at each
point in space, as mentioned in the Motivation. Conversely, if the
boundary condition (\ref{eq:BCp}) is not obeyed, the solution of
Eq.~(\ref{mu0}) yields $\cV[n_i(-\infty)] \ne \cV[n_i(+\infty)]$ for a
flat interface, that is, two macroscopic phases in thermodynamic
equilibrium are not in mechanical equilibrium, thus leading to a
contradiction.

Note that there are no separate constraints on individual
concentration gradients; such constraints could be thought of as some
matching conditions with regard to {\em partial} pressures of the
components, between the two phases. Only in a pure substance, does
Eq.~(\ref{eq:BCp}) place a hard constraint on the concentration
gradient. In any event, we observe that when the coefficients
$\kappa_{ij}$ experience a jump across the inter-phase boundary, so do
generally the concentration gradients and hence the quantities $\sum_j
\kappa_{ij} (d n_j/d r)$.  In the aforementioned mechanics analogy,
each quantity $\sum_j \kappa_{ij} (d n_j/d r)$ corresponds with a
momentum: $\prtl f/\prtl (d n_i/d r) = \sum_j \kappa_{ij} (d n_j/d
r)$.  The present results are a consequence of the N\"other theorem:
In the mechanics analogy, the concentration-dependent $\kappa_{ij}$'s
correspond to coordinate-dependent masses. For a Lagrangian that
depends explicitly on the coordinate but not on time, the energy is
still conserved, but not the momentum. We reiterate that within the
formal analogy between the Landau-Ginzburg density functional theory
and the Lagrange formulation of mechanics, Pascal's law corresponds to
energy conservation.


For concreteness, choose the dividing surface at the intersection of
the two paraboloids $\cV^{(\text{in})}$ and $\cV^{(\text{out})}$, see
Eq.~(\ref{Vpara}), corresponding to the phases inside and outside the
droplet respectively:
\begin{equation} \label{eq:BCdiv}
  \cV^{(\text{in})}\left( \{ n_i^\ddag \} \right) =
  \cV^{(\text{out})} \left( \{ n_i^\ddag \} \right),
\end{equation}
while assuming that the coefficients $\kappa_{ij}$ may experience a
discontinuity also at the very same surface.  Under these
circumstances, the boundary condition (\ref{eq:BCp}) simply states the
that the dynamic term of the free energy functional (\ref{Fmulti})
must be continuous across the dividing surface:
\begin{equation} \label{eq:BCp1} \sum_{ij}
  \left. \frac{\kappa_{ij}}{2} \frac{d n_i}{d r} \frac{d n_j}{d r}
  \right|_{R^-}^{R^+} = 0.
\end{equation}
Eqs.~(\ref{eq:BCconc}), (\ref{eq:BCp}), and (\ref{eq:BCdiv}) together
represent $M+2$ constraints.

While natural, the choice of boundary in Eq.~(\ref{eq:BCdiv}) is not
obligatory. If the boundary is chosen so that the bulk free energy
density is {\em discontinuous}, then the most general boundary
condition from Eq.~(\ref{eq:BCp}) must be used. If the discontinuity
in the coefficients $\kappa_{ij}$ takes place at concentration values
other than the dividing surface for the bulk free energy density,
there will be another dividing surface in real space. The present
methodology can be straightforwardly extended to such situations.

Now, the linear equations (\ref{mu0linear}), subject to the boundary
conditions (\ref{bc}) are solved by the following functions, except in
certain degenerate cases that can be handled straightforwardly:
\begin{equation} \label{yukawa} n_i = \left\{ \begin{array}{ll}
      n_i^{(\text{out},b)} + \sum_{jl} C_l^{(j)} e^{-k_l r}/r, &
      \hspace{3mm} r \ge R \\   \\
      n_i^{(\text{in},b)} + \sum_{jl} C_l^{(j)} \sinh(k_l r)/r, &
      \hspace{3mm} r < R,
    \end{array}
  \right.
\end{equation}
where the coefficients $C_l^{(j)}$ and parameters $k_l$ satisfy the
following equations (for each $i$ and phase):
\begin{equation} \sum_j^M C^{(j)} (\kappa_{ij} k^2 - m_{ij}) = 0.
\end{equation}
Since the characteristic equation $\text{Det}(\kappa_{ij} k^2 -
m_{ij}) = 0$ has $M$ roots, one can pick at most $M$ linearly
independent vectors $C^{(j)}$ (each of which has $M$
components). Consequently, the values of the coefficients at one
harmonic determines the coefficients for the rest of the
harmonics. Thus there are only $M$ independent coefficients
$C_l^{(j)}$ in each phase. Together with the radius $R$, which must be
determined self-consistently, there are $2M+1$ unknowns.

We thus have $(2M+1) - (M+2) = M-1$ unknowns to be determined, which
is exactly the dimensionality of the dividing surface in the $\{ n_i
\}$ space.  It is convenient to take the values of the transition
state concentrations $n_i^\ddagger$, $i = 1, \ldots, (M-1)$, as the
unknowns. The reason why the problem is so far underdetermined is that
we have solved for stationary-state configurations within individual
phases but not {\em at} the interface. Writing down equations at the
phase boundary would require one to separately specify the behavior of
the coefficients $\kappa_{ij}$ and the bulk free energy $\cV$ in that
region. A quick way to see that the number $(M-1)$ of undetermined
variables actually makes sense is to pretend that the components could
chemically interconvert at the interface. In the presence of such
interconversion, there would be imposed exactly $(M-1)$ constraints on
the concentrations $n_i^\ddagger$. (There are $M$ equations, $\dot
n_i^\ddagger = 0$, of which one is automatically satisfied because of
particle conservation.)  Instead of explicitly treating dynamics at
the interface, we note that since equations (\ref{mu0}) constitute a
free energy optimization problem, the steady state configuration can
be determined by minimizing the free energy function
(\ref{Fmulti})---as computed using the concentration profiles
(\ref{yukawa})---with respect to the undetermined quantities, i.e.,
the concentrations $n_i^\ddagger$, $i = 1, \ldots, (M-1)$ at the phase
boundary. Simply stated, these $(M-1)$ variables fully specify the
chemical composition at the interface.

The above methodology can be straightforwardly extended to a
non-stationary case, namely, a spherically symmetric droplet that is
evaporating or growing; some caveats will apply. According to the
above discussion, a droplet whose radius $R$ is greater or less than
its stationary, critical value will grow indefinitely and evaporate,
respectively. One may make a steady-state approximation, by which the
concentrations within the phases are assumed to equilibrate much
faster than the droplet growth/decay. Generically, such an
approximation is adequate when $\dot R/R \ll D_\text{diff}/R^2$, where
$D_\text{diff}$ is the diffusivity of the pertinent species. In this
approximation, one uses the steady-state version of Eq.~(\ref{nidot})
on either side of the dividing surface but not {\em at} the dividing
surface:
\begin{equation} \label{sstate} \nabla^2 \mu_i = 0, \hspace{5mm} r
  \gtrless R,
\end{equation}
where we have assumed that $\widetilde D_{ij} = \widetilde D_i
\delta_{ij}$ and the diffusivities $\widetilde D_i$ are
concentration-independent within each phase. The above equations are
supplemented by the general boundary conditions, c.f. Eq.~(\ref{bc}):
\begin{equation} \label{bcmu} \begin{array}{rcl} \mu_i (\br = \infty)
    &=& 0  \\  \\
    \bm \nabla \mu_i (\br = 0) &=& 0.
    \end{array}  
\end{equation}
At the phase boundary, the following conditions must be obeyed:
\begin{align}
  \mu_i(R^-) &= \mu_i(R^+) \label{eq:BCmu} \\
  \frac{1}{n_i^\ddag} \left.  \left(\widetilde{D}_i \frac{\prtl
        \mu_i}{\prtl r}\right)\right|_{R^-}^{R^+} &= \dot{R} \times
  \text{sign} \! \left[ n_i^{(\text{in}, b)} - n_i^{(\text{out}, b)}
  \right] \label{eq:Rdot}
\end{align}
The chemical potential must be continuous to prevent uncompensated
microscopic fluxes, hence Eq.~(\ref{eq:BCmu}). On the other hand,
Eq.~(\ref{eq:Rdot}) ensures particle conservation at the boundary: An
influx/outflow of the components at the phase boundary results in the
growth or evaporation of the droplet. For instance, in a pure
substance, the total particle influx, $4 \pi R^2 \widetilde{D} (\prtl
\mu/\prtl r)|_{R^-}^{R^+}$, must be equal to $4 \pi R^2 n^\ddag \dot
R$ for a liquid droplet surrounded by vapor but will be equal to $- 4
\pi R^2 n^\ddag \dot R$ for a bubble of vapor surrounded by
liquid. This is because a liquid droplet grows by collecting
particles, while a bubble grows by {\em giving up} particles. The
latter notion is reflected by the sign function in
Eqs.~(\ref{eq:Rdot}).  These equations, which apply to all $M$
components, represent $(M-1)$ constraints.
The rate $\dot R$ is deduced at the end of the calculation.

Eqs.~(\ref{sstate}), together with boundary conditions (\ref{bcmu})
and (\ref{eq:BCmu}), imply that each chemical potential has the
following functional form:
\begin{equation} \label{mur} \mu_i = \left\{ \begin{array}{ll}
      \mu_i^\ddagger, &r \le R \\ \\ \mu_i^\ddagger (R/r), &r > R.
    \end{array}
  \right.
\end{equation}
Thus compared with the original steady-state problem, there are $M$
new unknown quantities $\mu_i^\ddagger$ to be determined, while there
are only $(M-1)$ new constraints. This allows one to solve for the
concentration profiles for values of the droplet radius $R$ other than
the stationary one.

\begin{figure}[t]
  \includegraphics[width= \figurewidth]{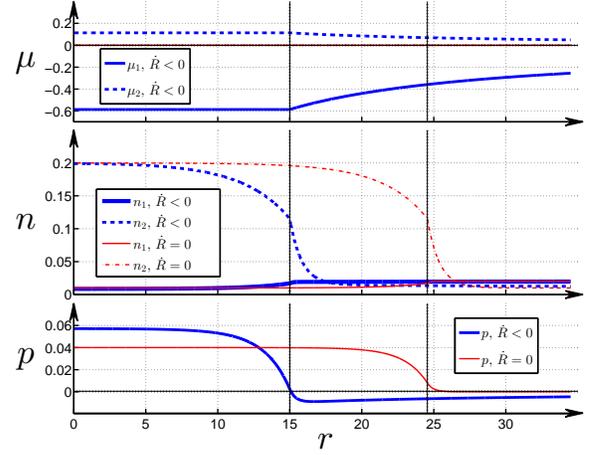}
  \caption{\label{munpr} The radial-coordinate dependences of the
    chemical potential $\mu$, concentrations $n_i$, and pressure $p$
    for a stationary ($\dot R = 0$, saddle-point) and non-stationary
    ($\dot R < 0$) configuration of a spherical droplet. Note the
    chemical potentials in the stationary case $\dot R = 0$ are equal
    to zero throughout: $\mu_1 = \mu_2 =0$.}
\end{figure}

\begin{figure}[t]
  \includegraphics[width= .8 \figurewidth]{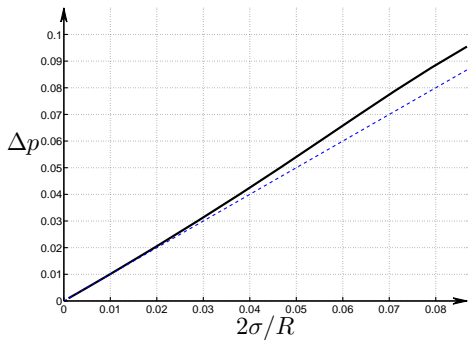}
  \caption{\label{Laplace} The pressure difference $\Delta p \equiv
    p(r=0)- p(r=\infty)$ plotted vs. the value $2 \sigma$ it would
    achieve in the thin interface limit, Eq.~(\ref{sigma}). Every
    point on both curves corresponds to a critical nucleus.}
\end{figure}

To avoid confusion, we point out that in a non-stationary case, one
must solve not the second-order equations (\ref{mu0}) but the
forth-order equations (\ref{nidot}), at $\dot n_i = 0$. Thus, like the
chemical potentials in Eq.~(\ref{mur}), the concentration profiles
will acquire $1/r$ contributions (for $r > R$) that reflect particle
exchange with the bulk. Although it involves solving transcendental
equations for $R$, in view of Eqs.~(\ref{yukawa}), the (numerical)
solution of the problem (\ref{nidot}) is straightforward and
computationally robust. In Fig.~\ref{munpr}, we graphically
demonstrate such a solution for the binary mixture from
Fig.~\ref{n1n2V} for both the stationary ($\dot R = 0$, $\mu_i = 0$)
and a non-stationary case ($\dot R < 0$, $\mu_i \ne 0$). The
corresponding $(n_1, n_2)$ trajectories connecting the droplet center
with the solution bulk are shown in Fig.~\ref{n1n2V}.  As remarked
earlier, one expects that in steady state, the pressure is spatially
uniform for a flat interface, while increasing monotonically toward
the origin for a droplet with non-zero curvature. We observe in
Fig.~\ref{munpr} that the converse is also true: For a non-critical
droplet, the pressure no longer changes monotonically with the radial
coordinate. In any event, the Laplace formula works reasonably well,
despite the interface being thick, see Fig.~\ref{Laplace}.

\begin{figure}[t]
  \includegraphics[width= 0.9 \figurewidth]{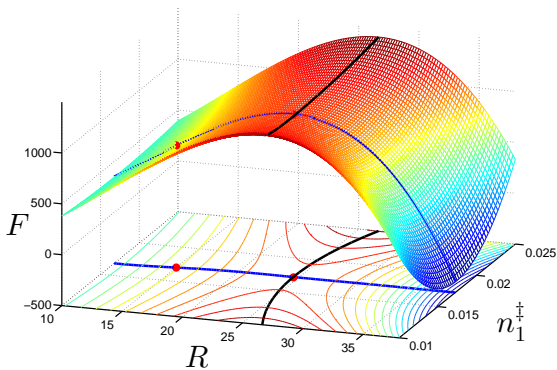}
  \caption{\label{Rn1daggerF} The free energy surface of the droplet
    as a function of the radius $R$ of the phase boundary and the
    concentration $n_1^\ddagger$ of component 1. The dots indicate the
    two droplet configurations from Fig.~\ref{munpr}. The line along
    the ridge of the surface corresponds to stationary points $\dot R
    = 0$; of these points, only the lowest one is a true solution of
    Eq.~(\ref{mu0}). The other line is the steepest descent line for a
    droplet prepared at the saddle point.}
\end{figure}

In Fig.~\ref{Rn1daggerF}, we show the droplet free energy
(\ref{Fmulti}) as a function of $R$ and $n^\ddagger_1$. In computing
this free energy for non-stationary cases $\dot R \ne 0$, the $1/r$
contributions were omitted. These contributions are associated with
the bulk, not the droplet itself, and, appropriately, lead to
divergent contributions to the free energy (\ref{Fmulti}). As already
discussed, $(M-1)$ concentrations $n^\ddagger_i$ are artificially not
dynamical in the treatment because we do not explicitly solve
Eqs.~(\ref{mu0}) at the phase boundary. This means that the $\dot R =
0$ points form a $(M-1)$-dimensional surface in the $(R,
n^\ddagger_i)$ space.  When $M=2$, this surface corresponds to a line;
the latter line can be seen to run through a ``ridge'' on the $F(R,
n^\ddagger_1)$ surface. The actual saddle point corresponds to the
lowest point on the ridge. If prepared exactly on the ridge but away
from the saddle point, the system will quickly relax to the saddle
point. The characteristic time scale is given by $R^2/\nu$, where $\nu
\equiv \eta/\rho$ is the kinematic viscosity, $\eta$ viscosity, and
$\rho$ mass density of the fluid. The kinematic viscosity, which has
dimensions of the diffusivity, significantly exceeds the diffusivity
itself at viscosities in question, i.e., $\eta > 10^{-2}$~Ps. This can
be seen straightforwardly by comparing $\eta/\rho$ and the
Stokes-Einstein $k_B T/(6 \pi \eta a)$, where $a$ is a molecular size.

If prepared away from the ridge, the droplet will quickly relax along
the $n^\ddagger_i$ coordinates while also moving, relatively slowly,
along the $R$ coordinate. The latter evolution will lead to droplet
growth or evaporation depending on which side of the ridge the system
was initially prepared.  The relaxation in the $n^\ddagger_i$
coordinates corresponds to changes in the composition at the droplet
interface. Such mixing processes involve hydrodynamic flows and are
expected during phase ordering, as discussed early on by
Siggia~\cite{PhysRevA.20.595} and later, within a
renormalization-group framework, by Bray.~\cite{Bray} While operating
primarily along the phase boundaries for small volume fractions of the
minority phase, such modes affect phase ripening when the latter
volume fraction is sufficiently high.  Last but not least, we note
that according to Fig.~\ref{Rn1daggerF}, the chemical composition at
the boundary depends only weakly on the droplet radius.

Note that the matching conditions (\ref{eq:BCp}) and (\ref{eq:BCmu})
between the minority and majority phase are formally identical to
those between phases that are in macroscopic equilibrium, as
elucidated by Gibbs a long time ago. These matching conditions cover
the chemical potentials of the individual species, pressure, and
temperature. In the case of macroscopic coexistence, for each
additional chemical species there could be, in principle, another
phase in coexistence with those phases already present. In the present
context, the number of phases is limited to two, by construction. The
extra, ``mixing'' degrees of freedom stemming from increasing the
number of components result in an increase of the dimensionality of
the dividing surface in the $\{ n_i \}$ space, between the minority
and majority phases.

\section{Amphiphilic Adsorbents at the Interface}
\label{amphi}

We now consider a situation where no pressure ``conservation'' takes
place. In mechanics, the energy is not conserved when the Lagrangian
explicitly depends on time. Likewise, in the present problem, the full
free energy density must explicitly depend on the coordinate:
\begin{equation} \label{Vfull} \cV = \cV_\text{n} + \sum_i n_i
  \cVext^{(i)}(\br),
\end{equation}
where, by construction, the quantity $\cVext^{(i)}(\br)$ is an
external potential acting on species $i$, $\prtl \cVext^{(i)}/\prtl
n_j = 0$, while $\cV_\text{n}$ does not explicitly depend on the
coordinate: $\prtl \cV_\text{n}/\prtl \bm r = 0$. More general forms
of the coordinate-dependent portion of the full free energy density
$\cV$ can be considered. For non-zero $\cVext^{(i)}(\br)$,
Eqs.~(\ref{mu}) can be written out as follows:
\begin{align} \label{mu0ext} \sum_j \bm \nabla (\kappa_{ij} \bm \nabla
  n_j)& - \frac{1}{2} \sum_{lm} \frac{\prtl \kappa_{lm}}{\prtl n_i}
  (\bm \nabla n_l \bm \nabla n_m) \nonumber \\ &= \frac{\prtl
    \cV_\text{n}}{\prtl n_i} + \cVext^{(i)}(\br) - \mu_i(\br),
\end{align}
while Eq.~(\ref{12}) becomes
\begin{eqnarray} \label{12ext} \sum_{ij} \left. \frac{\kappa_{ij}}{2}
    \frac{d n_i}{d r} \frac{d n_j}{d r} \right|_1^2 &-& \cV_\text{n} |_1^2 =
  - (D-1)\int_1^2 \frac{dr}{r} \sum_{ij} \kappa_{ij} \frac{d n_i}{d
    r}\frac{d n_j}{d r} \nonumber \\ &+& \sum_i \int_1^2 d n_i
  [\cVext^{(i)}(r) - \mu_i(r)],
\end{eqnarray}
where a spherically-symmetric geometry is understood.  Equations
(\ref{p}) through (\ref{psph}) can be rewritten for the specific form
of the full free energy density from (\ref{Vfull}). And so for a flat
interface,
\begin{equation} \label{pflat1} p = -\cV_\text{n} + \sum_{ij}
  \frac{\kappa_{ij}}{2} \frac{d n_i}{d x} \frac{d n_j}{d x} - \sum_i
  n_i [\cVext^{(i)}(x) - \mu_i(x)].
\end{equation}
Thus, according to Eqs.~(\ref{12ext}) and (\ref{pflat1}), we conclude
that the quantity on the l.h.s. of Eq.~(\ref{12ext}) does not
correspond to a pressure difference, nor is it generally conserved in
the presence of an external potential, even for a flat interface in
complete equilibrium, $\mu_i = 0$.

\begin{figure}[t]
  \includegraphics[width= 0.8 \figurewidth]{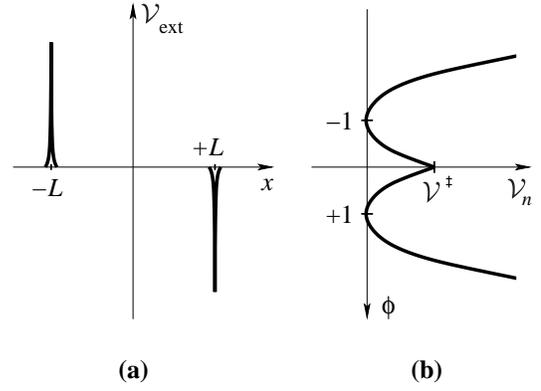}
  \caption{\label{VamFig} {\bf (a)} Schematic of the one-particle
    potential $\cVext$ from Eq.~(\ref{Vamphi}) for $\epsilon > 0$.
    {\bf (b)} The concentration-dependent portion of the bulk free
    energy density $\cV_\text{n}$ from Eq.~(\ref{Vfull}), employed in
    the present work to analyze the amphiphilic action. By
    Eq.~(\ref{Vpara}), $\cV^\ddagger = m/2$.}
\end{figure}

Still, Eq.~(\ref{pflat1}) implies that pressure changes continuously
with the coordinate for a smooth external potential. In the spirit of
the preceding Section, we next consider a {\em singular} potential so
as to cause the pressure to experience a discontinuity.  For the sake
of argument, let us consider a flat interface in a one-component
system. We choose the following form for the coordinate-dependent
portion of the bulk free energy:
\begin{equation} \label{Vamphi} \cVext(x) = -\epsilon \delta(x - L) +
  \epsilon \delta(x + L),
\end{equation}
see sketch in Fig.~\ref{VamFig}(a). The above potential can be viewed
as an extreme limit of a smoother potential that has an attractive
minimum centered at $x = +L$ and a repulsive ``hump'' centered at $x =
- L$.

By construction, the shapes of the attractive and repulsive parts are
exactly the same; one may think of the potential in Eq.~(\ref{Vamphi})
as originating from a ``dipole'' of sorts.  The latter potential can
be thought of a stationary, externally-imposed field, but can be
equally well thought of as resulting from adding an amphiphilic
species---or detergent---to the system. Indeed, suppose we are not
concerned with the total density in a binary solution but only with
the {\em partial} quantity of the components. The partial quantity can
be described using a single order parameter $\phi$. This way, the
derivative $\prtl \cV/\prtl \phi$ is related to the actual chemical
potentials of the components in a linear fashion. For convenience, one
may choose the minima of the free energy density $\cV$ to be at $\phi
= \pm 1$, see Fig.~\ref{VamFig}(b).  The potential (\ref{Vamphi}) can
be thought of as a field acting on linear molecules so that their two
tails prefer to stay in the phases that are relatively rich in the two
components of the mixture respectively. Specifically, for the bulk
free energy density in Fig.~\ref{VamFig}(b), the l.h.s. and
r.h.s. peaks of $\cVext$ in Fig.~\ref{VamFig}(a) serve as a repulsive
bump and attractive minimum, respectively, in the $\phi > 0$
phase. The situation is reversed in the $\phi < 0$ phase. The quantity
$\epsilon$, by construction, reflects the surface concentration of the
amphiphiles but also their solvation free energy. We will call the
quantity $\epsilon$ the {\em activity} for brevity. To avoid confusion
we note that this is {\em not} the standard definition of the activity
of a component in a solution. (Note the quantity $\epsilon$ could have
either sign.)
 
What are the appropriate boundary conditions for a singular potential
of the type from Eq.~(\ref{Vamphi})? A definitive answer to this
question can be obtained, if the coefficient $\kappa$ is a continuous
function of the concentration near $x=\pm L$. (The situation to the
contrary will be discussed shortly.)  Because the potential $\cVext$
is infinitely narrow, $\kappa$ can be regarded as a constant on the
lengthscale of the spatial variation of $\cVext$. In the mechanics
analogy, we now have a situation where the {\em energy} is no longer
conserved: Each $\delta$-function peak in Eq.~(\ref{Vamphi})
corresponds to an infinitely rapid ``kick.'' The resulting increment
in the momentum is given by the time integral of the kick, by Newton's
third law. This notion can be implemented in the present context by
integrating Eq.~(\ref{mu0ext}) in terms of $x$ from $L^-$ to $L^+$
(and likewise for the l.h.s. peak from Eq.~(\ref{Vamphi})). One gets,
as a result:
\begin{equation} \label{BCmomentum} \left. \kappa \frac{d \phi}{d x}
  \right|_{\pm L^-}^{\pm L^+} = \mp \epsilon.
\end{equation}
Note that using Eq.~(\ref{12ext}) would {\em not} yield unambiguous
boundary conditions because computing the integral $\int d \phi \,
\cVext = \int dx (d \phi /dx) \, \cVext$ requires the knowledge of the
derivative $d \phi /dx$ {\em exactly} at the point of
singularity. However, the quantity $d \phi/dx$ experiences a
discontinuity at the singularity and thus is not well-defined
there. Incidentally, we point out that a problem in which a
$\delta$-function-like potential is placed exactly at the phase
boundary is generally ill-defined mathematically in the sense that one
cannot place a constraint on the concentration gradient of the type
from Eq.~(\ref{eq:BCp}) or (\ref{BCmomentum}). To appreciate this, one
may invoke the mechanics analogy once again and think of the
singularities in the bulk free energy density $\cV_\text{n}$ and the
one-particle potential $\cVext$ not as confined to an infinitely
narrow region of the coordinate and time respectively, but, instead,
as smeared over {\em finite} ranges of the respective variables. At
the same time, the values of the momenta specified on the two sides of
the original singularity should be now thought of as specified at
infinity. Presented in this form, the setup of the problem is entirely
analogous to how one thinks of scattering events.  In one-dimensional
space, one can evaluate the particle's momentum long after such a
scattering event without having to solve detailed equations of motion
at intermediate times, if either energy or momentum are conserved (or
both), but not otherwise. We thus conclude that a limit when a
discontinuity in $\cV$ or its derivative and a $\delta$-function-like
potential $\cVext$ exactly coincide in space, is ill-defined. This is
not a real issue in practice, since the distance (coordinate-wise)
between the two singularities can be made arbitrarily small; we shall
see this shortly.

\begin{figure}[t]
  \includegraphics[width= 0.9 \figurewidth]{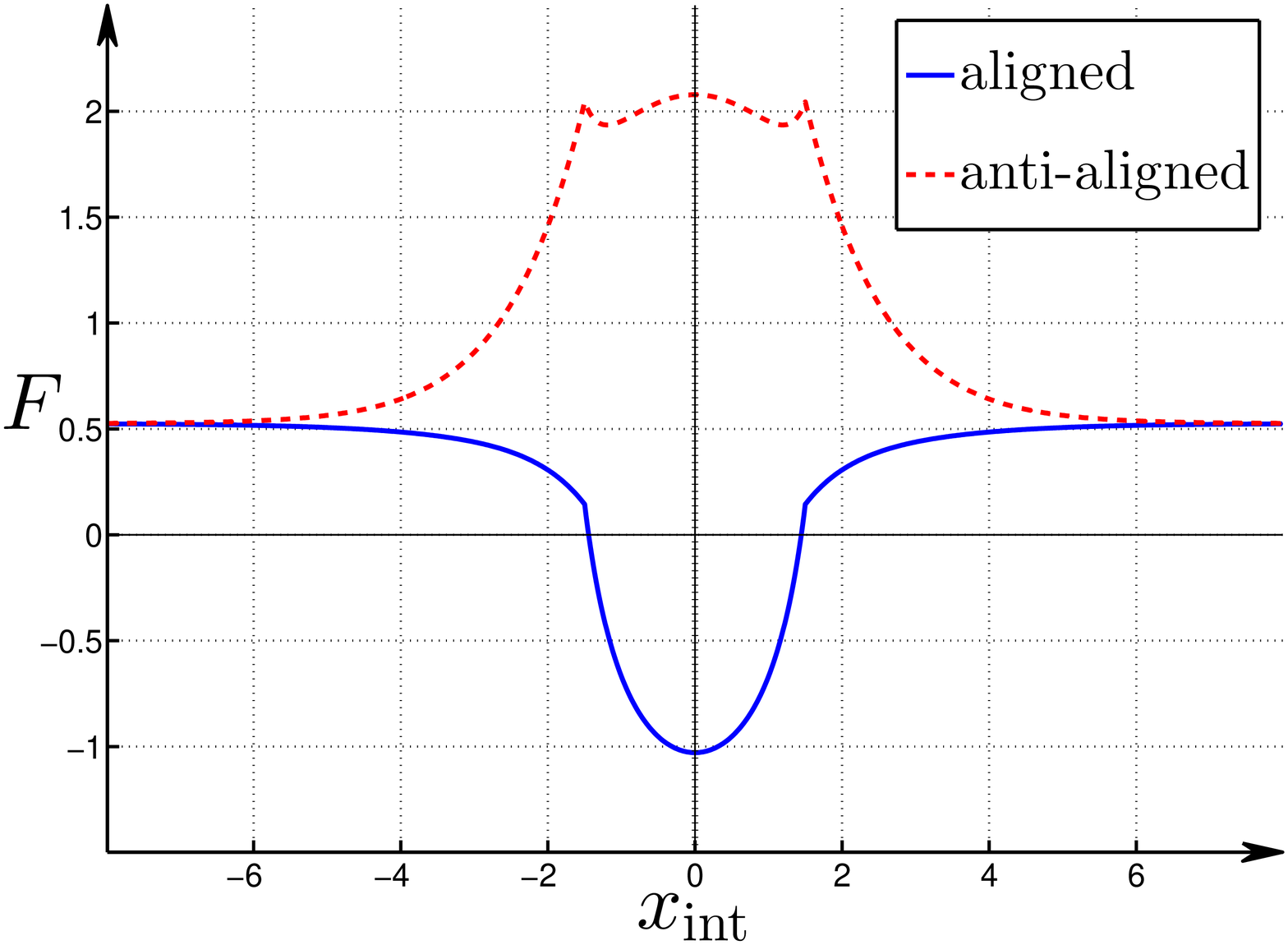}
  \caption{\label{Sigmaxint} The free energy of the interface in the
    presence of potential $\cVext$ from Eq.~(\ref{Vamphi}), plotted as
    a function of the distance $x_\text{int}$ between the phase
    boundary and the midpoint of the potential. Here, $L = 1.5$,
    $\epsilon = +1$ bottom term, $\epsilon = -1$ top term.}
\end{figure}

To illustrate the above discussion we find the density profile for a
flat interface in the presence of external potential (\ref{Vamphi}),
subject to boundary conditions: 
\begin{align} \label{bcamphi}
  \phi(+\infty) &= 1 \\ \phi(-\infty) &= -1. \label{bcamphi1}
\end{align}
For clarity, we assume $\kappa = \text{const}$.  It is obvious from
the symmetry of the problem that the optimal position of the phase
boundary $x_\sint$, defined as $\phi(x_\sint) = 0$, is exactly at the
origin. (The subscript ``int'' refers to ``interface.'') The origin,
we remind, is at the midpoint of the ``dipole''-like potential in
Eq.~(\ref{Vamphi}) by construction.  This notion can be made even more
explicit by plotting the free energy of the system as a function of
$x_\sint$. Since $x_\sint \ne 0$ configurations are non-stationary, we
must solve the full problem (\ref{nidot}) under such circumstances. As
in the preceding Section, we instead solve a steady-state problem in
which $\nabla^2 \mu = 0$ is everywhere except at $x = \pm L$ and the
phase boundary. In 1D, this equation is solved by a linear function,
leaving us with the following option, up to an additive constant:
\begin{equation} \label{mu1D}
  \mu(x) = \left\{ \begin{array}{lc} 0, & x < x_1, \,  x_3 \le x  \\
      \mu_0 (x-x_1)/(x_2-x_1), & x_1 \le x < x_2 \\ \mu_0
      (x_3 - x)/(x_3-x_2), & x_2 \le x < x_3,
  \end{array} \right.
\end{equation}
where $x_1$, $x_2$, and $x_3$ ($x_1 < x_2 < x_3$) are assigned to the
locations of the three singularities in the ascending order. The
presence of the sloped portion of the chemical potential in
Eq.~(\ref{mu1D}) reflects that away from equilibrium, the material
must be exchanged within the $(x_1, x_3)$ interval---but not between
the bulk phases!---in order to relax toward equilibrium. The
appearance of a new unknown $\mu_0$ allows us to regard the distance
$x_\sint$ as a free parameter. The resulting interface free energy per
unit area, as a function of $x_\sint$, is shown in
Fig.~\ref{Sigmaxint}. In this Figure, we show the free energy in two
specific situations, where the pinning potential is ``aligned'' with
the boundary conditions (\ref{bcamphi})-(\ref{bcamphi1}), $\epsilon >
0$, and anti-aligned $\epsilon < 0$, respectively. While the
attractive minimum and overall repulsion near $x_\sint = 0$ for the
aligned and anti-aligned configurations respectively are expected, it
is less obvious that there should be some structure to the repulsive
maximum in the upper term.  Here and in the rest of the Section, we
use units such that the length $\sqrt{\kappa/m}$ and the surface
tension coefficient of a standalone interface, which is $\sqrt{\kappa
  m}$ by Eq.~(\ref{sigma1}), are both set to unity.

\begin{figure}[t]
  \includegraphics[width= \figurewidth]{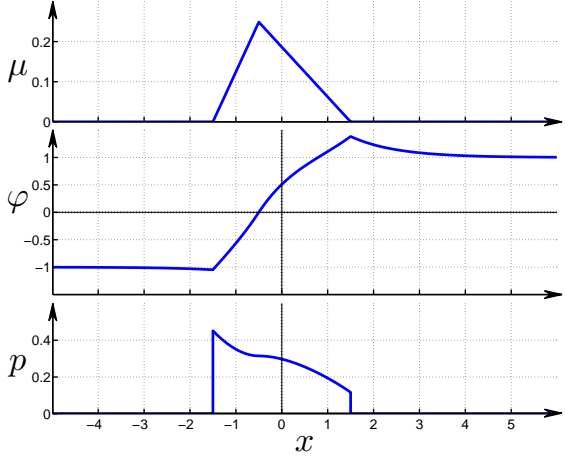}
  \caption{\label{mupsipin} The coordinate dependences of the chemical
    potential $\mu$, order parameter $\phi$, and pressure $p$ for a
    specific configuration corresponding to $x_\sint = - 0.5$ of the
    bottom (bonding), term from Fig.~\ref{Sigmaxint}. $L =
    1.5$. Hereby, the phase boundary is located between the peaks of
    the potential $\cVext$ from Eq.~(\ref{Vamphi}). The
    $\delta$-function contributions to pressure are not
    shown. $\epsilon =1$.}
\end{figure}

\begin{figure}[t]
  \includegraphics[width= \figurewidth]{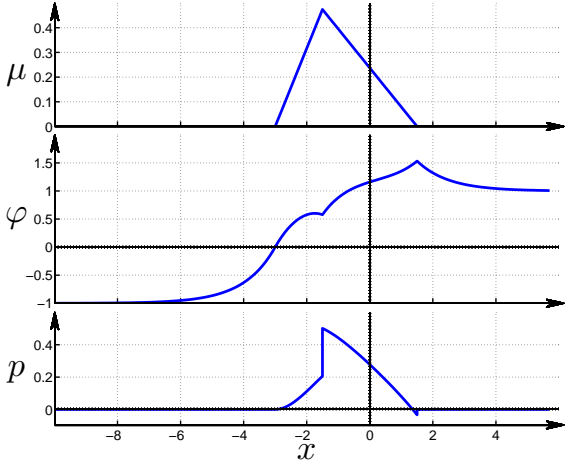}
  \caption{\label{mupsipout} The coordinate dependences of the
    chemical potential $\mu$, order parameter $\phi$, and pressure $p$
    for a specific configuration corresponding to $x_\sint = - 3$ of
    the bottom (bonding), term from Fig.~\ref{Sigmaxint}. $L =
    1.5$. The phase boundary is to the left of both peaks of the
    potential $\cVext$ from Eq.~(\ref{Vamphi}),
    c.f. Fig.~\ref{mupsipin}. $\epsilon =1$.}
\end{figure}

We now illustrate two specific configurations of the interface for
$x_\sint \ne 0$, pertaining to the bottom term in
Fig.~\ref{Sigmaxint}. These configurations correspond to two distinct
mutual arrangements of the phase boundary and the potential, see
Figs.~\ref{mupsipin} and \ref{mupsipout}, where the phase boundary is
located, respectively, between the singularities of the pinning
potential $\cVext$ and outside (to the left of) those singularities.

Note that in the symmetric cases, $x_\sint = 0$ (not shown), which are
both optima of the free energy, the chemical potential is zero
throughout. Under these circumstances, the pressure is now a conserved
quantity and thus must be uniform in the $-L < x < L$ interval, by
Eqs.~(\ref{pflat}).  (Recall, the potential $\cVext$ is zero in that
interval.) Within the $-L < x < L$ interval, the pressure is positive
and negative for the aligned and anti-aligned configurations,
respectively.

\begin{figure}[t]
  \includegraphics[width= 0.9 \figurewidth]{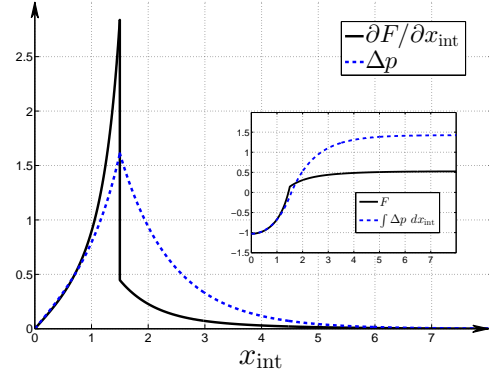}
  \caption{\label{dFdxintdpxint} Solid line: The coordinate derivative
    of the free energy of the interface in the presence of potential
    $\cVext$ from Eq.~(\ref{Vamphi}), plotted as a function of the
    distance $x_\text{int}$ between the phase boundary and the
    midpoint of the potential. Dashed line: Difference between the
    pressure discontinuities at the ends of the amphiphile, also as a
    function of $x_\text{int}$.}
\end{figure}

It is interesting to ask how how big a portion of the free energy
$F(x_\sint)$ in Fig.~\ref{Sigmaxint} accounts for the work needed to
move the amphiphile; the remainder is used to redistribute the liquid
itself. In answering this question, we show in
Fig.~\ref{dFdxintdpxint} the derivative of the free energy
$F(x_\sint)$ with respect to the location $x_\sint$ and the difference
between the pressure discontinuities at points $x = \pm L$, both as
functions of $x_\sint$. The integrals of both quantities w.r.t. to the
coordinate are shown in the inset.  One observes that the mechanical
work expended in bringing the amphiphile from the bulk is
substantially offset by the free energy cost of redistribution of the
solvent near the amphiphile.

\begin{figure}[t]
  \includegraphics[width= 0.9 \figurewidth]{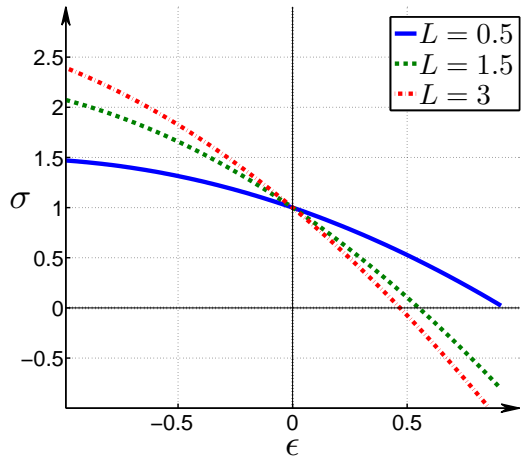}
  \caption{\label{SigmaEpsilon} Dependence of the free energy of the
    interface on the activity $\epsilon$ of the amphiphile, for three
    distinct values of $L$. All curves correspond to the stable
    minimum in Fig.~\ref{Sigmaxint}, $x_\sint = 0$, bottom term.}
\end{figure}

The rest of the Section focuses exclusively on the equilibrium
configuration, which corresponds to the minimum in the bottom,
``bonding'' term in Fig.~\ref{Sigmaxint}, located at $x_\sint = 0$. Of
greatest interest is the degree of renormalization of the surface
tension as a function of the activity $\epsilon$ of the amphiphile,
which is illustrated in Fig.~\ref{SigmaEpsilon}, for three specific
values of the amphiphile half-length $L$. Note that for any value of
$L$, there are critical values of the activity beyond which the
interface tension is exactly compensated by the ``solvation'' energy
of the amphiphile leading to a vanishing of the effective surface
tension. This can be seen explicitly in Fig.~\ref{SigmaEpsilon}. We
note that for sufficiently large and {\em negative} values of
$\epsilon$, the curves in Fig.~\ref{SigmaEpsilon} bend down and cross
the origin. This is because in the (somewhat trivial) limit $\epsilon
\to \pm \infty$, the intrinsic energy scale of the interface becomes
negligible and the response of the heterophase system can be
approximated by that of a pure phase.

The value of the surface tension $\sigma$ cannot be negative, of
course. As $\sigma$ becomes sufficiently small---but still
positive---several things could happen after more amphiphile is added
to the solution, depending on circumstances.  If the interface is not
allowed to break up into smaller pieces, it will spontaneously begin
to distort so as to increase in area and become able to accommodate
additional detergent. This is a realization of the Krafft
point.~\cite{MyersSurfactantBook} The distortion of the interface will
occur at a {\em positive} value of $\sigma$, even if relatively small,
because of the non-zero vibrational entropy of the interface. On the
other hand, if equilibrated---which could be facilitated by
stirring---the system can lower its free energy by breaking the
original interface into smaller pieces thus leading to the formation
of a {\em suspension} of the minority phase covered with the
detergent. This is, of course, how soap works.

\section{Interface Wetting in Glassy Liquids}
\label{wettingSection}

As already remarked, the potential (\ref{Vamphi}) can be viewed either
as that due to an amphiphilic molecule or, equally well, as an
externally imposed potential. This potential pins the interface,
Fig.~\ref{Sigmaxint}, but also down-renormalizes its tension,
Fig.~\ref{SigmaEpsilon}. Down-renormalization of the surface tension
is of central significance in the random field Ising
model~\cite{Villain} (RFIM) and in the context of the free energy
landscape and activated transport in glassy liquids.~\cite{LW_ARPC,
  L_AP} In contrast with conventional phases of matter, as originally
envisioned by Gibbs, glassy liquids are characterized by a complex
free energy landscape, where the number of free energy minima scales
exponentially with the system size $N$: $e^{s_c N}$, where $s_c$ is
the configurational entropy per particle.~\cite{MCT1, KTW, XW} This
multiplicity of minima reveals itself directly as the excess liquid
entropy relative to the corresponding crystal.  According to the
random first order transition (RFOT) theory,~\cite{LW_ARPC} liquid
transport is realized via activated transitions between the free
energy minima. The minima are deep enough that the lifetime of the
structures corresponding to each minimum greatly exceeds the
vibrational time.  The activated transport on the landscape proceeds
via nucleation-like events; the free energy cost for reconfiguring a
compact region of size $N$ reads:
\begin{equation} \label{FN} F(N) = - T s_c N + \gamma \sqrt{N}. 
\end{equation}
The bulk-driving force for reconfiguration $(- T s_c N)$ is due to the
exponential multiplicity of free energy minima.  The surface tension
portion $\gamma \sqrt{N}$ scales with the droplet radius $R \propto
N^{1/D}$ as $R^{D/2}$, not the conventional $R^{D-1}$. This
corresponds to an effective surface tension coefficient that decreases
with the radius as $R^{D/2}/R^{D-1} = R^{1-D/2}$ and thus vanishes for
macroscopic interfaces in any number of spatial dimensions higher than
$D=2$. A detailed argument~\cite{Villain, KTW, LRactivated} shows that
this renormalization can be thought of as resulting from a distortion
of the original, relatively thin interface so as to locally optimize
free energy.~\cite{KTW} The latter fluctuations are transiently
frozen-in and obey the usual Gaussian statistics. The
down-renormalization of the mismatch penalty results in a lowering of
the barrier for activated transport. Quantitative estimates~\cite{XW,
  RL_sigma0, StevensonW, LRactivated, RWLbarrier} for the barrier are
in good agreement with observation, without using adjustable
parameters. As a result of the distortions, the interface becomes
fractal in shape, with an effective width scaling with the droplet
radius itself.  An analogy between free energy fluctuations in a
glassy liquid and the random on-site field of the random-field Ising
model (RFIM) was originally pointed out and exploited by Kirkpatrick,
Thirumalai, and Wolynes,~\cite{KTW} while a procedure to map a
specific liquid-model onto the RFIM has been developed by Stevenson et
al.~\cite{stevenson:194505}

In a related picture put forth by Bouchaud and
Biroli,~\cite{BouchaudBiroli} each long-lived compact structure can be
thought of as fitting its environment better than a generic aperiodic
arrangement of particles within the corresponding region. Hereby, the
partition function of the region, with a fixed environment (up to
vibration), consists of two contributions: that of an essentially
unique contribution of the well-fitting structure and that of the rest
of the configurations:
\begin{equation} \label{ZN} Z(N) = e^{-\beta (-\gamma N^x)} + e^{s_c
    N}.
\end{equation}
In the Bouchaud-Biroli (BB) argument, the exponent $x$ at the
stabilization free energy $(-\gamma N^x)$ is not specified, except
that it must be less than one. The argument is agnostic as to the
detailed mechanism of escape from the long-lived state.

The pictures leading to Eqs.~(\ref{FN}) and (\ref{ZN}) are
complementary~\cite{LW_aging} and can be thought of as differing with
regard to the free energy reference. The former picture starts from a
stabilized state and considers the free energy cost of escaping from
such a stable state, which is incurred because two randomly chosen
distinct structures do not mutually fit. In the BB picture, one starts
from a high free energy state as resulting from bringing in contact
two dissimilar structures; the strain is uniformly high in such a
state. One then picks through alternative structures within a compact
region of some size $N$ until the best match is found that lowers the
free energy to the greatest extent.  The two pictures lead to the same
size $N^*$ of the cooperatively rearranging region:
\begin{equation} \gamma \sqrt{N^*} = T s_c N^*,
\end{equation}
while the exponent $x$ from Eq.~(\ref{ZN}) may be fixed at $1/2$, see
a detailed discussion by Lubchenko and Rabochiy~\cite{LRactivated}
(and also a recent review~\cite{L_AP}). The latter authors have argued
the term $\gamma \sqrt{N}$ can be thought of as the magnitude of a
typical fluctuation of the Gibbs free energy at size $N$. Consistent
with the large breadth of the distorted interface, the mismatch
penalty between the region and its environment is distributed over the
whole region, in contrast with the DFT-based picture from
Section~\ref{CNT}, in which the penalty is accrued largely over a
narrow shell.

In the context of the preceding Section, one may view the random field
of the RFIM (and, hence, the frozen-in fluctuations of the RFOT) as a
collection of stationed amphiphilic molecules that are distributed
uniformly in space.  The molecules are randomly oriented and sized.
Indeed, because the on-site field is zero on average, for each site
with a field oriented in a particular direction, there will be a site
with a field equal in magnitude and opposite in direction.

It is easy to see that beyond a certain threshold value of the
magnitude of the on-site field, an originally smooth interface will
distort spontaneously. As discussed in Section~\ref{amphi}, if the
activity of the amphiphile exceeds a certain value, the total amount
of interfacial regions begins to increase. This results either in a
bigger number of droplets or in a distortion of the interface, if the
number of droplets cannot change for some reason. (The two options are
not as distinct as one might think, see below.) Suppose we start with
a relatively thin interface confined to a region of size $N$
particles; the interface separates two dissimilar aperiodic
structures.  We thus set our reference state in the Bouchaud-Biroli
way. In the presence of the ``amphiphiles,'' the boundary will distort
some.  Since the amphiphiles are distributed uniformly in space, the
interface will come in contact with a new set of the amphiphiles and
distort some more. The interface will continue distorting until the
interfacial material covers all the amphiphiles contained within the
region.

\begin{figure}[t]
  \includegraphics[width= 0.7 \figurewidth]{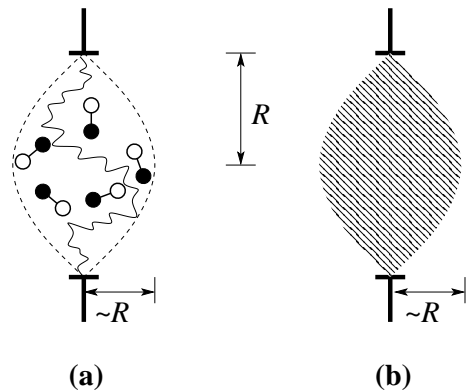}
  \caption{\label{Villain} {\bf (a)} Cartoon depicting a distorted
    interface subjected to uniformly seeded amphiphilic
    molecules. {\bf (b)} A renormalized view of panel (a), in which
    the amphiphiles are effectively accounted for through a thicker,
    softened interface.}
\end{figure}

One can view the distorted interface as having the same {\em bare}
surface tension coefficient as a standalone undistorted interface but
stretched and covering more amphiphiles, see Fig.~\ref{Villain}(a).
In this view, the decreased {\em effective} surface tension is due to
the solvation energy of the amphiphiles that are in contact with the
interface, despite the greatly increased area of the distorted
interface.  The distortion takes place in a continuous range of
wavelengths. The process of distortion can be broken up into a
sequence of elemental steps, each of which occurs in a narrow
wavelength range $[r, r+dr]$.~\cite{Villain, KTW, LRactivated, L_AP}
This sequence constitutes a continuous renormalization-group
transformation.  The end result of this transformation, up to
wavelength $R$, is an effective interface that is not stretched but is
thicker and {\em softer}, see sketch in Fig.~\ref{Villain}(b).  The
renormalized interface covers fewer amphiphiles, because some of the
amphiphiles are already effectively accounted for through the softened
tension.  The renormalization-group argument is
standard;~\cite{Villain, KTW} it was revisited
recently~\cite{LRactivated} and will not be repeated here.  The role
of the on-site field is now played by the local value of the
amphiphile activity $\epsilon$.  Lubchenko and
Rabochiy~\cite{LRactivated} used an RG line of thought to argue that
given a sufficiently large value of $R$, surface renormalization
occurs for any positive value of $\epsilon$, no matter how small. The
basic reason is that the stabilization due to the distortion scales as
$\epsilon (\delta N)^{1/2}$, where $\delta N$ is the number of
amphiphiles ``swept'' during an individual step in the renormalization
procedure. The penalty for stretching of the interface scales with a
higher power of $\delta N$ and thus always inferior to $\sqrt{\delta
  N}$ at sufficiently small $\delta N$, so long as $\epsilon
>0$.~\cite{LRactivated}


The scaling of the eventual amount of renormalization with the region
size can be deduced with little effort by noting that the extent of
renormalization scales approximately linearly with the total activity
of the amphiphilic molecules residing within the boundary, see
Fig.~\ref{SigmaEpsilon}. Since the orientations of the latter
molecules are random, the total activity is on average zero, but could
be either positive or negative for a specific configuration of the
interface. The magnitude of such total activity fluctuation scales as
the usual $\sqrt{N}$, by the law of large numbers.  The latter law
applies when the correlation length $\sqrt{\kappa/m}$ is not too
large, which is indeed the case.~\cite{RL_sigma0} Negative
fluctuations will be more likely than the positive ones, according to
Boltzmann. These negative fluctuations will lead to a stabilization of
the region in question, the amount of stabilization scaling as
$\sqrt{N}$, as just said. The latter stabilization corresponds with
the $-\gamma N^x$ term in Eq.~(\ref{ZN}), at $x=1/2$, and is
consistent with the picture advanced by the RFOT theory.


As follows from the above discussion, a macroscopically large flat
interface will continue to distort indefinitely, in the presence of
uniformly seeded amphiphiles ($R \to \infty$ limit in
Fig~\ref{Villain}).  This is why the effective surface tension
coefficient of a flat interface {\em vanishes} in the RFOT theory. One
may say that there is no conventional surface tension between
alternative aperiodic minima of free energy in glassy liquids, in the
sense of the classical nucleation theory.  It turns out that the main
formula of this article, Eq.~(\ref{p}), provides an additional
perspective on this fascinating fact, if combined with earlier ideas
of Xia and Wolynes.~\cite{XW} The latter authors pictured the surface
tension renormalization as a ``wetting'' of the interface between two
distinct meanfield free energy minima of the glassy liquid by
configurations that are representative of {\em other} meanfield
minima.~\cite{XW}

\begin{figure}[t]
  \includegraphics[width= 0.9 \figurewidth]{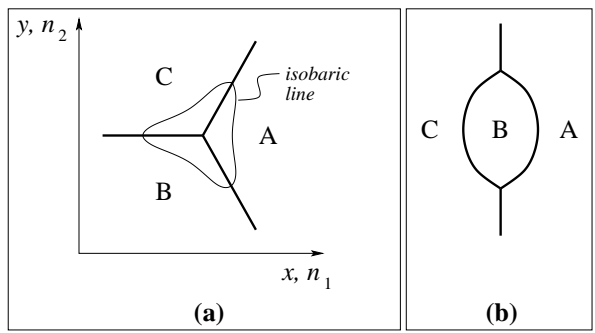}
  \caption{\label{wetting} {\bf (a)} Three co-existing phases A, B,
    and C in contact. The horizontal axis denotes both the Cartesian
    axis $x$ and, roughly, the direction of the $\bm \nabla n_1$
    gradient. Likewise, the vertical axis denotes both the Cartesian
    axis $y$ and the direction of the $\bm \nabla n_2$ gradient. In
    this arrangement, isobaric lines are not too different from the
    $\cV = \text{const}$ lines, by Eq.~(\ref{p}).  {\bf (b)} A droplet
    of phase B sandwiched between phases A and C. Regions of
    three-phase coexistence can be approximated as in the (a) panel.}
\end{figure}

In considering interface wetting, let us use a bulk free energy
density that does not explicitly depend on the coordinate: $\prtl
\cV/\prtl \bm r = 0$.  To set the stage, we first consider the
geometry in which three nearly flat, intersecting interfaces separate
some three distinct liquid phases, call them A, B, and C. The three
interfaces intersect along the same straight line, see
Fig.~\ref{wetting}(a). This geometry can also be thought of as
approximating a small portion of an oblate bubble of phase B residing
between two distinct phases A and C, see Fig.~\ref{wetting}(b).
Formally, to realize such a three-phase fluid coexistence, the liquid
must consist of at least two components, by the Gibbs phase rule. By
construction---and for future convenience---suppose phase A is
relatively rich in component 1 and intermediate with regard to the
content of component 2. Phase B is poor in both components, and C is
poor in component 1 and rich in component 2, see
Fig.~\ref{wetting}(a).  The isobaric surfaces are roughly similar to
the $\cV = \text{const}$ surfaces, but certainly not to the $n_i =
\text{const}$ surfaces. The former surfaces are represented by closed
contours centered generically at the three-phase boundary. The $n_i =
\text{const}$ surfaces, in contrast, are approximately parallel to the
$x$ and $y$ axes on Fig.~\ref{wetting}(a).

Because the external normal to an isobaric surface is no longer
collinear with the concentration gradients, the first term under the
sum in Eq.~(\ref{p}) is significantly reduced compared with a
spherically symmetric situation, much more so than the second
term. This is particularly obvious if one sets the off-diagonal
elements $\kappa_{ij}$, $i \ne j$ to zero. (A quick thought shows that
the contribution of the off-diagonal terms in the geometry from
Fig.~\ref{wetting}(a) is relatively small even if the off-diagonal
$\kappa_{ij}$ are non-zero, because the $\bm \nabla n_i$ gradients are
mutually, approximately orthogonal.) To estimate the reduction in the
first term, assume the gradients of each concentration $n_i$ preserve
their directions as one goes around an isobaric surface. Meanwhile,
the external normal $\bm e$ performs a full rotation per each closed
path around the line. Under these circumstances, the average $\la (\bm
e \bm \nabla n_i) ( \bm e \bm \nabla n_j) \ra \simeq (1/D) \la (\bm
\nabla n_i \bm \nabla n_j) \ra$ in $D$ spatial dimensions. Indeed, the
rotational average $\la e_i e_j \ra = \delta_{ij}(1/D)$, where $e_i$
is the $i$-th component of the unit vector $\bm e$. This is because
$\la e_i e_j \ra \propto \delta_{ij}$ and $\sum_i e_i e_i =1$.

We thus observe that in the geometry of Fig.~\ref{wetting}(a), $D=2$,
the two terms under the sum in Eq.~(\ref{p}) largely compensate each
other. The latter equation thus yields that the region in which the
three phases coexist is at {\em negative} pressure, relative to the
bulk. This is expected and constitutes a mechanically-stable geometry:
The line where the three phases meet can be thought of as being
suspended and pulled on by the two-phase boundaries. Likewise, a point
in 3D space where four distinct liquid phases meet, would be suspended
on the lower-order phase boundaries and would be even at yet lower
pressure, since the first term under the sum in Eq.~(\ref{p}) would
get only smaller with $D$.

Imagine now a spherical droplet of one phase surrounded by another
phase and suppose that the boundary is wetted by a large number of
alternative phases similarly to how the A-C interface is wetted by
phase B in Fig.~\ref{wetting}(b). Imagine that new interfaces
appearing during wetting are further wetted and so on, down to the
correlation lengthscale $\sqrt{\kappa/m}$ of the Landau-Ginzburg
theory for an individual phase. According to the above argument, the
pressure must {\em decrease} as one moves inward from the bulk toward
the center of the droplet. Now, because there are no phase boundaries
to suspend the low pressure region on, the latter region would {\em
  collapse}.  Alternatively, this situation could be thought of a
corresponding to a negative surface tension coefficient, by
Eqs.~(\ref{vdW}) and (\ref{sigma}). At their face value, the above
notions are simply a consequence of the Gibbs rule: Given a finite
number of components, only a finite number of phases is allowed and so
the unlimited wetting described above is impossible.

Yet in glassy liquids, the number of possible ``phases,'' which
correspond with distinct free energy minima, grows exponentially with
volume and is thus effectively unlimited. Hence we conclude that the
mismatch penalty between distinct free energy minima in a glassy
liquid cannot be defined in the conventional way. This notion is
consistent with our earlier discussion of the surface tension
renormalization in the presence of a random field, but it also fleshes
out the ``wetting'' perspective on the renormalization of the mismatch
penalty.

In applying the present result to glassy liquids, an important caveat
must be made. Strictly speaking, the functional (\ref{Fmulti}) applies
to {\em fluids}, in which the shear modulus is zero. In arriving at
Eq.~(\ref{FN}), one adopts the view that the ``minority phase'' is an
equilibrated liquid, by construction. The environment, on the other
hand, represents a mechanically stable (aperiodic) solid, on the
pertinent timescale. A solid can, in fact, support a compact region of
negative pressure. This notion does not invalidate the present
analysis so long as the glassy liquid is equilibrated on the
laboratory timescale and so the reconfigurations occur without a
volume change. Still, there may be some residual stresses in the
environment due to local inhomogeneity in stress. Such inhomogeneities
would be present even in a periodic crystal, let alone an aperiodic
one.~\cite{PhysRevLett.97.055501, 0953-8984-26-1-015007, Marruzzo2013}
Already the spatially-inhomogeneous bulk free energy from
Eq.~(\ref{Vfull}) captures, in part, long-range correlations that may
arise as a result. Indeed, the field from an elastic ``dipole'' made
of two point-like forces can be emulated using the free energy
functional (\ref{Fmulti}) by making the coefficients $m_{ij}$ from
Eq.~(\ref{Vpara}) sufficiently small. A full treatment, however, must
include off-diagonal entries of the strain-tensor. Such treatments are
available at the strict mean-field~\cite{BL_6Spin} and higher,
Onsager-cavity~\cite{BLelast} levels.

\section{Summary}
\label{summary}

This work analyzes several phenomena in the area of phase transitions
from a {\em mechanical} vantage point. It takes advantage of a formal
analogy between the Lagrangian formalism in classical mechanics and
the Landau-Ginzburg description of macroscopic phase coexistence, in
which the analog of time is the spatial coordinate that traverses the
phase boundary. Pressure, which is the partial derivative of the free
energy with respect to that spatial coordinate, is formally analogous
to the mechanical energy, which is the partial derivative of the
action with respect to time. This analogy allows one to generalize the
old---and surprisingly little known---results of van der Waals on
phase coexistence to multi-component mixtures and certain
off-equilibrium situations.  While automatically yielding that
Pascal's law must be obeyed during macroscopic coexistence, the
approach can be extended straightforwardly to spherical geometries of
the minority phase and, in some cases, to other geometries. The so
obtained mechanical perspective on phase coexistence allows one to
make progress in several problems that are seemingly disparate yet
turn out to be connected.

In the first problem, of interest to mesoscopic phases in liquid
solutions, we establish that certain singular forms of the bulk free
energy density are amenable to essentially analytical solution in
individual phases. To patch the solutions at the phase boundary, where
the bulk free energy density is singular, one must require that
pressure be a continuous function of the coordinate. We
straightforwardly obtain a saddle-like free energy profile for a
nucleus of the minority phase. The coordinate transverse to the
nucleus size corresponds to the chemical composition at the
interface. The coordinate is interesting in that its dynamics do not
involve particle exchange with the bulk; it thus must operate along
the phase boundary, consistent with earlier, hydrodynamics-based
analysis by Siggia.~\cite{PhysRevA.20.595, Bray} The chemical
composition at the interface is seen to depend only weakly on the
droplet radius.

The opposite situation of the {\em violation} of Pascal's law leads us
to a model that has a double use: Surface tension renormalization in
the presence of a pinning potential or in the presence of amphiphiles.
Simple relations between the amount of renormalization and the
activity of the amphiphile are established. 

In the two setups above, Pascal's law is obeyed and violated
respectively. Although seemingly distinct, the two situations turn out
to be two faces of the same coin in the final problem analyzed in this
work.  Here we argue that the complicated problem of
down-renormalization of the mismatch penalty between alternative
aperiodic structures in glassy liquids can be thought about in a
relatively simple fashion: On the one hand, the renormalization can be
viewed as a distortion of the interface separating two dissimilar
structures in the presence of uniformly seeded, randomly oriented
amphiphiles.  On the other hand, the renormalization of the mismatch
can be viewed as extensive wetting of the interface by yet other
aperiodic structures. In both views, Pascal's law is violated, yet in
the latter view, it is violated only transiently since the bulk free
energy does not explicitly depend on the coordinate. This reflects
that the disorder in glassy liquids is self-generated, not quenched.
Both views indicate that the mismatch between alternative structures
cannot be described using conventional notions of the density
functional theory even in the thick interface limit. Yet the
amphiphile view, which is rooted in those same notions, allows a way
out of this seeming impasse: The renormalization can be thought of an
essentially bulk effect due to a random field (as stemming from the
amphiphiles), which is zero on average. When it happens to be non-zero
in any specific region, this field scales at most as the square root
of the region size, thus resulting in a square root scaling of the
mismatch penalty with the region size. This scaling is consistent with
the RFOT theory.

{\em Acknowledgment.} The authors thank Peter G. Wolynes and Peter
G. Vekilov for inspiring conversations. They gratefully acknowledge
support by the National Science Foundation (MCB-1244568, CHE-0956127)
and the Welch Foundation Grant E-1765.


%

\end{document}